\documentclass[journal]{IEEEtran}
\IEEEoverridecommandlockouts
\usepackage{lettrine}
\usepackage{cite}
\usepackage{amsmath,amssymb,amsfonts}
\usepackage{algorithm}
\usepackage{algorithmic}
\usepackage{bm}
\usepackage{graphicx}
\usepackage{subfigure}
\usepackage{tabularx}
\usepackage{textcomp}
\usepackage{xcolor}
\usepackage{stfloats}
\usepackage{multirow}
\def\BibTeX{{\rm B\kern-.05em{\sc i\kern-.025em b}\kern-.08em
		T\kern-.1667em\lower.7ex\hbox{E}\kern-.125emX}}

\floatname{algorithm}{\bf{Algorithm}}  
\renewcommand{\algorithmicrequire}{\textbf{Input:}}  
\renewcommand{\algorithmicensure}{\textbf{Output:}}

\begin{document}
	\title{Element-Grouping Strategy for Intelligent Reflecting Surface:  Performance Analysis and Algorithm Optimization}
	\author{Shengsheng~Zhang, Taotao~Ji, Meng~Hua,~\IEEEmembership{Member,~IEEE,} Yongming~Huang,~\IEEEmembership{Fellow,~IEEE,} and~Luxi~Yang,~\IEEEmembership{Senior~Member,~IEEE}%
		\IEEEcompsocitemizethanks{\IEEEcompsocthanksitem Shengsheng Zhang, Taotao Ji, Yongming Huang, and Luxi Yang are with the School of Information Science and Engineering, the National Mobile Communications Research Laboratory, and the Frontiers Science Center for Mobile Information Communication and Security, Southeast University, Nanjing 210096, China, and also with the Pervasive Communications Center, Purple Mountain Laboratories, Nanjing 211111, China (e-mail: zhangshengsheng@seu.edu.cn; jitaotao@seu.edu.cn; huangym@seu.edu.cn; lxyang@seu.edu.cn).
			\IEEEcompsocthanksitem Meng Hua is with the Department of Electrical and Electronic Engineering, Imperial College London, London SW7 2AZ, UK (e-mail: m.hua@imperial.ac.uk).
	}}
	
	\maketitle
	
	\begin{abstract}
		As a revolutionary paradigm for intelligently controlling wireless channels, intelligent reflecting surface (IRS) has emerged as a promising technology for future sixth-generation (6G) wireless communications. While IRS-aided communication systems can achieve attractive high channel gains, existing schemes require plenty of IRS elements to mitigate the ``multiplicative fading'' effect in cascaded channels, leading to high complexity for real-time beamforming and high signaling overhead for channel estimation. In this paper, the concept of sustainable intelligent element-grouping IRS (IEG-IRS) is proposed to overcome those fundamental bottlenecks. Specifically, based on the statistical channel state information (S-CSI), the proposed grouping strategy intelligently pre-divide the IEG-IRS elements into multiple groups based on the beam-domain grouping method, with each group sharing the common reflection coefficient and being optimized in real time using the instantaneous channel state information (I-CSI). Then, we further analyze the asymptotic performance of the IEG-IRS to reveal the substantial capacity gain in an extremely large-scale IRS (XL-IRS) aided single-user single-input single-output (SU-SISO) system. In particular, when a line-of-sight (LoS) component exists, it demonstrates that the combined cascaded link can be considered as a ``deterministic virtual LoS'' channel, resulting in a sustainable squared array gain achieved by the IEG-IRS. Finally, we formulate a weighted-sum-rate (WSR) maximization problem for an IEG-IRS-aided multiuser multiple-input single-output (MU-MISO) system and a two-stage algorithm for optimizing the beam-domain grouping strategy and the multi-user active-passive beamforming is proposed. Simulation results validate the superiority of our proposed two-stage algorithm in low pilot overhead conditions and show that in the context of an XL-IRS aided MU-MISO system, the proposed IEG-IRS can achieve a significant WSR gain, thus overcoming this performance drawback associated with high complexity and signaling overhead.
		
	\end{abstract}
	
	
	\section{Introduction}
	\IEEEPARstart{O} {ver} the past decades, a variety of wireless technologies have been proposed and thoroughly investigated to increase system capacity, while wireless channels are always considered uncontrollable \cite{9998527}. Intelligent reflecting surface (IRS) has been recently proposed as a promising technology for proactively modifying wireless environment via highly reconfigurable and intelligent signal reflection \cite{8910627}. Specifically, an IRS is generally designed as a planar array composed of a large number of reconfigurable passive elements that independently reflect electromagnetic signals in a desired manner to collaboratively reconfigure the signal propagation \cite{9140329}. Benefit from the advantages of high array gain, low energy consumption, and low hardware cost \cite{10496514,10025617,9335617}, IRS promises to achieve greater system capacity \cite{10016300}, broader coverage area \cite{10268052}, and higher spectral efficiency \cite{9309152}. Therefore, IRS is expected to realize a sustainable wireless communication evolution in future sixth-generation (6G) network \cite{8811733}.
	
	In the most prior works on IRS-aided wireless communication system, it is generally assumed that the channel state information (CSI) of all links involved is perfectly known, see \cite{10070785,10536035,10143420}, etc. Particularly, in an $N$-element IRS aided single-user single-input single-output (SU-SISO) system, the asymptotic channel gain achieved is proportional to $N^2$, which can scale down the transmit power without compromising the user signal-to-noise ratio (SNR) \cite{8811733}. Benefit from this ``squared gain" behind, it is expected to utilize a larger number of IRS elements to achieve greater performance improvements. Nevertheless, given that the IRS's elements are passive, there are no radio-frequency (RF) chains available and the accurate CSI associated with the IRS can be only obtained via the base station (BS)/access point (AP) in fact, which leads to the high signaling overhead introduced by $N$ pilots for channel estimation and the high complexity of $\mathcal{O}(N^2)$ for real-time beamforming \cite{9090356,9839429}. Besides, the introduction of the IRS gives rise to the ``multiplicative fading" effect, whereby the equivalent path loss of the cascaded transmitter-IRS-receiver link is the product of the path loss of between the transmitter-IRS link and IRS-receiver link, and is thousands of times larger than that of the direct link \cite{9306896}. Therefore, an extremely large-scale IRS (XL-IRS) is required to mitigate the ``multiplicative fading" effect. Many previous methods for optimizing reflection coefficients are not scalable for XL-IRS as the number of optimization variables becomes unaffordably large. By integrating reflection-type amplifiers into their reflecting elements, the active IRS has the ability to actively reflect signals with amplification, thereby reducing the number of IRS elements \cite{9998527}. However, this requires higher power consumption as well as more complex hardware architecture. As a consequence, to advance the practicability of the XL-IRS in future 6G wireless networks, the fundamental performance bottlenecks caused by the high complexity and signaling overhead are urgent to be broken.
	
	To overcome those fundamental performance limitations, in this paper, a new IRS architecture called intelligent element-grouping IRS (IEG-IRS) is proposed for wireless communication system. Some existing works have revealed the potential advantages of the grouping IRS. In \cite{8937491,9039554}, the idea of IRS elements grouping is first proposed, in which the adjacent elements are divided into the common group and the training overhead and design complexity are significantly reduced by adjacent-element grouping IRS (AEG-IRS). Then, to analyze the performance of AEG-IRS, the authors in \cite{10018442} derived a closed-form upper bound of the adjacent-element grouping strategy dependent on the statistical CSI (S-CSI). Subsequently, the authors in \cite{9729398} present the optimal adjacent-element grouping strategy by analyzing the impacts of statistical channel parameters, group size, and the distance between adjacent IRS elements. Different from the same variation pattern being employed by all groups \cite{9039554,9677923,10018442}, the authors in \cite{10286054} propose the use of different variation patterns for the groups to avoid very low beam gains at specific directional angles of a channel. In contrast to those adjacent-element grouping strategies, the elements of the proposed IEG-IRS in each group are distributed flexibly at arbitrary locations within the IRS. Furthermore, even as the number of IRS elements in each group increases, our proposed grouping strategy continues to deliver a significant performance improvement. Specifically, inspired by AEG-IRS \cite{8937491,9039554,9677923,10018442,9729398,10286054}, an intelligent element-grouping strategy is designed to improve channel array gain in low pilot overhead conditions. The S-CSI, which incorporates the deterministic line-of-sight (LoS) component, is employed to intelligently pre-group IRS elements. Subsequently, the combined reflection-coefficient vector (RCV) of the IEG-IRS is optimized using the instantaneous CSI (I-CSI) after grouping. Different from the general ungrouped IRS (U-IRS) that reflects the incident signals of each element individually, the IEG-IRS can adjust the combined RCV in group by the proposed grouping strategy, wherein the elements in each group share the common phase shift. Especially, in SU-SISO system, the components of the combined cascaded channel in each group reflected by IEG-IRS are coherently added with the close phase at the receiver. Therefore, in low pilot overhead conditions, IEG-RIS is capable of compensating for the large ``multiplicative fading'' of reflected links while significantly reducing computational complexity for real-time beamforming. 
	
	In this paper, we propose the concept of IEG-IRS and focus on the signal model, asymptotic performance analysis and two-stage algorithm. Specifically, our main contributions are summarized as below:
	
	\begin{itemize}
		\item We propose a novel grouping strategy and introduce the signal model for the IEG-IRS within the framework of an XL-IRS aided communication system. The channel array gain available is enhanced by the IEG-IRS. In particular, as the number of IRS elements increases, the communication performance continues to enhance without incurring additional pilot overhead from the transceiver system.
		\item Based on the introduced signal model, we analyze the asymptotic performance gain of an IEG-IRS aided SU-SISO system with an extremely large number of IRS elements over the Rician channel. Notably, when a LoS component is available, the asymptotic performance gain achieved is proportional to the square of the total number of IRS elements in low pilot overhead conditions. Furthermore, in this context, it demonstrates that the combined cascaded channel reflected by the IEG-IRS can be approximated as a ``deterministic virtual LoS'' link.
		\item To evaluate the performance of IEG-IRS in typical communication systems, we formulate a weighted-sum-rate (WSR) maximization problem for an IEG-IRS aided multi-user multi-input single-output (MU-MISO) system. Then, by exploiting the quadratic transform for fractional programming (FP) \cite{8314727} and alternating direction method of multipliers \cite{8186925}, a two-stage algorithm for optimizing the beam-domain grouping strategy and multi-user active-passive beamforming is proposed to solve the formulated problem.
		\item Simulation results validate the effectiveness of the proposed two-stage algorithm for optimizing the beam-domain grouping strategy and multi-user active-passive beamforming, demonstrating a substantial performance improvement compared to other benchmark schemes in low pilot overhead conditions.
	\end{itemize}
	
	\subsection{Organization \& Notations}
	The rest of this paper is organized as follows. Section II introduces the signal models of U-IRS and IEG-IRS, respectively. In Section III, the asymptotic performance gain and performance loss of the IEG-IRS are analyzed. In Section IV, a WSR maximization problem is formulated for an IEG-IRS aided MU-MISO system, and a two-stage algorithm design for optimizing the beam-domain grouping strategy and the multi-user active-passive beamforming is proposed to solve the formulated problem in Section IV. In Section V, simulation results are presented to evaluate the performance of the proposed designs in typical communication scenarios. Finally, we conclude the paper in Section VI.
	
	\textit{Notations:} $\mathbb{C}$, $\mathbb{R}$, $\mathbb{R}_ +$ and $\mathbb{Z}$ denote the sets of complex, real, positive real and integer numbers, respectively; $\left [ \cdot  \right ]^{-1}$, $\left [ \cdot  \right ]^* $, $\left [ \cdot  \right ]^T$, and $\left [ \cdot  \right ]^H$ denote the inverse, conjugate, transpose, and Hermitian transpose operations, respectively; $\left | \cdot \right | $, $\left \| \cdot \right \| $ and $\left \| \cdot \right \|_1$ denotes the amplitude, Euclidean norm and $L_1$-norm of the argument, respectively; $\mathrm{diag}\left ( \cdot \right )$ denotes the diagonalization operation, and $\mathrm{arg}\left(\cdot\right)$ denotes the phase of the argument; $\odot$ denotes the Hadamard product; $\mathbb{E}\left ( \cdot \right )$ denotes the statistical expectation; $\Re\left ( \cdot \right )$ denotes the real part of the argument; $\mathcal{O}$ denotes the big-O notation; $\mathcal{CN}\big(\boldsymbol{\mu}, \boldsymbol{\Sigma}\big)$ denotes the complex multivariate Gaussian distribution with mean $\boldsymbol{\mu}$ and variance $\boldsymbol{\Sigma}$; $\mathbf{I}$ and $\mathbf{0}$ are an identity matrix and an all-zero matrix, respectively, with appropriate dimensions.

	\section{U-IRS and IEG-IRS}
	In this section, we present two different architectures of IRSs. First, in Subsection II-A, we review U-IRS and point out its performance limitation imposed by the high complexity and pilot overhead. Subsequently, in Subsection II-B, we propose a novel concept of IEG-IRSs to overcome those fundamental limitations. Finally, in Subsection II-C, we consider an IEG-IRS aided MU-MISO communication system.
	
	\begin{figure}[htbp]
		\centerline{\includegraphics[width=\columnwidth]{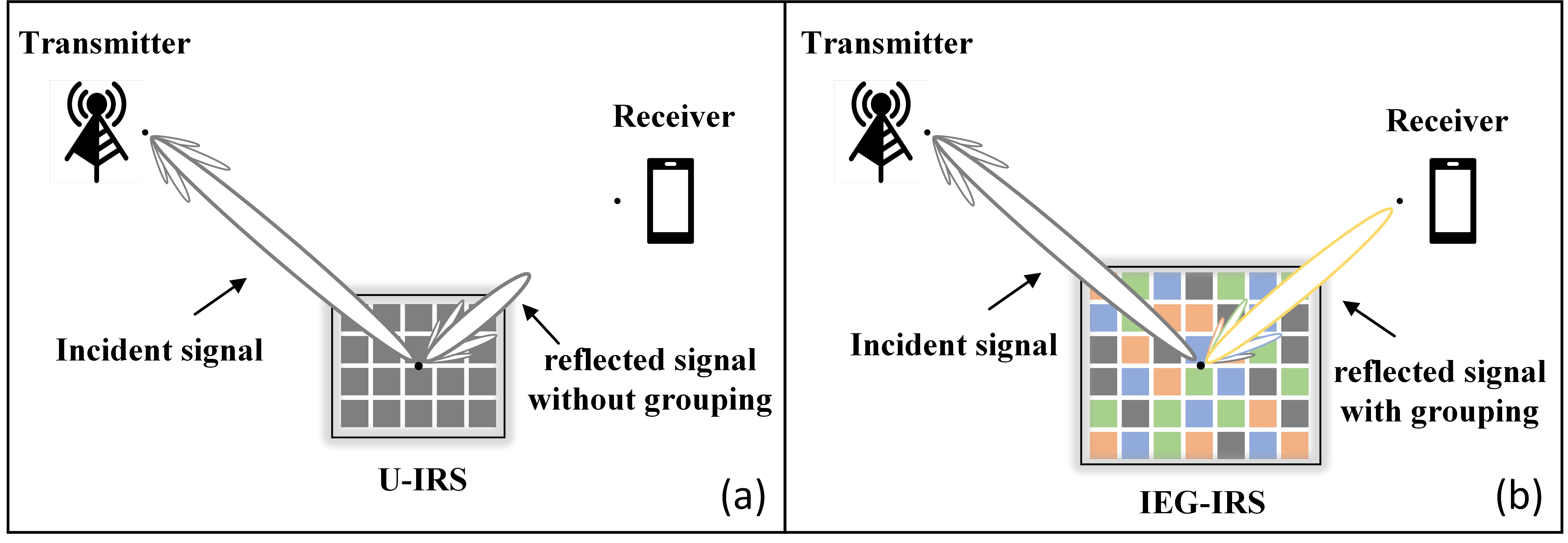}}
		\caption{An illustration of signal models of a U-IRS (a) and an IEG-IRS (b).}
		\label{fig:1}
	\end{figure}
	
	\subsection{U-IRS}
	Currently, the IRSs widely studied in most existing works are ungrouped \cite{9998527, 8811733, 9090356}. Specifically, a U-IRS comprises $Q$ reflecting elements, closely spaced at half-wavelength intervals, each of which is capable of reflecting the incident signal with an independent and controllable phase shift, as shown in Fig. 1 (a) \cite{cui2014coding}. Let $\check{\mathbf{\Phi}} = \mathrm{diag}\left\{ {\check{\mathbf{v}}^H} \right\} \in \mathbb{C}^{Q \times Q}$ denote the diagonal reflection-coefficient matrix of the U-IRS, where $\check{\mathbf{v}} = \big[ {\check{v}_1},\cdots,{\check{v}_q},\cdots ,{\check{v}_Q} \big]^H \in \mathbb{C}^{Q \times 1}$ and $\check{v}_q = {{e^{j{\check{\theta} _q}}}}$ are the RCV and the phase shift of the $q$-th IRS element, respectively. Then, the signal model of an $Q$-element U-IRS is given as \cite{8796365}:
	\begin{equation}
		\label{eq:1}
		\check{\mathbf{r}} = \check{\mathbf{\Phi}} \check{\mathbf{x}},
	\end{equation}
	where $\check{\mathbf{x}} \in \mathbb{C}^{Q \times 1}$ and $\check{\mathbf{r}} \in \mathbb{C}^{Q \times 1}$ denote the incident and reflected signals, respectively. 
	
	Then, by properly adjusting the RCV $\check{\mathbf{v}}$ to manipulate the signal $\check{\mathbf{r}}$ reflected by the $Q$ elements to coherently add with the same phase at the receiver, a high array gain can be obtained. The ability expected to achieve high gain has recently attracted lots of research interests in U-IRS \cite{8930608,9326394,9110849,9115725}.
	
	However, in practice, the high array gain expected is challenging to achieve \cite{9998527}. The is because the signal path loss of the reflected link, i.e., the cascaded BS-IRS-user link, is several orders higher than that of the direct link due to the ``multiplicative fading'' \cite{9306896}. As a consequence, the signal model requires an XL-IRS equipped with plenty of IRS elements to compensate for the ``multiplicative fading'' effect. In addition, given that the reflecting elements are passive, the IRS lacks the capability to estimate the accurate channel state information (CSI). Most existing channel estimation methods only obtain the CSI of the cascaded BS-IRS-User channel \cite{9866003}. However, as the dimensions of the XL-IRS elements increase significantly, they lead to prohibitively high computational complexity and signaling overhead, making scalable optimization of the XL-IRS quite challenging \cite{9306896}. In order to improve the practicality of the XL-IRS in general wireless communication, two fundamental performance bottlenecks, namely high complexity and signalling overhead, are urgent to be addressed.
	
	\subsection{IEG-IRS}
	In this paper, a novel IRS scheme called IEG-IRS is proposed as a promising solution to overcome the fundamental performance bottlenecks caused by the XL-IRS. Different from the U-IRS that just reflects the incident signals of each element individually, as illustrated in Fig. 1 (b), the IEG-IRS can adjust the combined RCV in group by the proposed grouping strategy to pre-divide the $N$ XL-IRS elements into $Q$ groups, wherein the elements in each group share the common phase shift. Here, a constraint is imposed that each group must contain at least one XL-IRS element, and each XL-IRS element is assigned exclusively to a specific group. Therefore, let $\mathbf{G} \in \mathbb{Z}^{Q \times N}$ denote the proposed grouping strategy matrix for an $N$-element and $Q$-group IEG-IRS, then the grouping problem can be formulated as
	\begin{subequations}
		\label{eq:2}
		\begin{align}
			{{\rm{Find}}}{\quad}&\mathbf{G}, \nonumber\\
			{\mathrm{s.t.}}{\quad}&{{G_{q,n}} \in \left\{ {0,1} \right\},} \label{eq:2A}\\
			{}&{\sum\limits_{q = 1}^Q {{G_{q,n}}}  = 1,} \label{eq:2B}\\
			{}&{\sum\limits_{n = 1}^N {{G_{q,n}}}  \ge 1,} \label{eq:2C}
		\end{align}
	\end{subequations}
	where $G_{q,n}=1$ is the $n$-th XL-IRS element assigned to the group $q$, while $G_{q,n}=0$ indicates that no grouping operation has been performed. 
	
	For an $N$-element IEG-IRS, there exist $\frac{1}{Q!}  {\textstyle \sum_{q=0}^{Q}}\left ( -1 \right )^q\binom{Q}{q}\left ( Q-q \right )^N    $ different grouping strategies, which is exceedingly large as Q increases, rendering the corresponding active-passive precoding optimization infeasible. In Section III-A, the idea of the grouping optimization strategy is initially proposed for an IEG-IRS aided SU-SISO system by the S-CSI. Subsequently, in Section IV-B, the aforementioned idea of the grouping strategy optimization is employed in an IEG-IRS aided MU-MISO system. 
	
	Let $\hat{\mathbf{\Phi}} \buildrel\textstyle.\over= \mathrm{diag}\left\{ {\hat{\mathbf{v}}^H \mathbf{G}} \right\} \in \mathbb{C}^{N \times N}$ denote the combined diagonal reflection-coefficient matrix for the IEG-IRS, where $\hat{\mathbf{v}} =\left [
		\hat{v}_1,\dots,\hat{v}_q,\dots,\hat{v}_Q\right ]^H \in \mathbb{C}^{Q \times 1}$ denotes the combined RCV after the grouping operation and $\hat{v}_q=e^{j{\hat{\theta} _q}}$ is the phase shift of the $q$-th group. Then, the signal model of an $N$-element IEG-IRS is given as
	\begin{equation}
		\label{eq:3}
		\hat{\mathbf{r}} = \hat{\mathbf{\Phi}} \hat{\mathbf{x}},
	\end{equation}
	where $\hat{\mathbf{x}} \in \mathbb{C}^{N \times 1}$ and $\hat{\mathbf{r}} \in \mathbb{C}^{N \times 1}$ denote the incident signal and the signal reflected by the IEG-IRS, respectively.
	
	By properly adjusting the combined RCV $\hat{\mathbf{v}}$, the signal $\hat{\mathbf{r}}$ reflected by the IEG-IRS coherently adds with the close phase at the receiver, a high array gain can be obtained. Compared with \eqref{eq:1}, the grouping strategy of the IEG-IRS intuitively reduces the equivalent dimensions of XL-IRS elements, thereby significantly reducing high computational complexity and pilot overhead. This implies that the combined dimensions of IEG-IRS elements are not directly correlated with the total number of the XL-IRS elements denoted by $N$, but rather with the number of the groups denoted by $Q$.
	
	\subsection{IEG-IRS Aided MU-MISO System}
	\begin{figure}[htbp]
		\centering
		\includegraphics[width=0.45\textwidth]{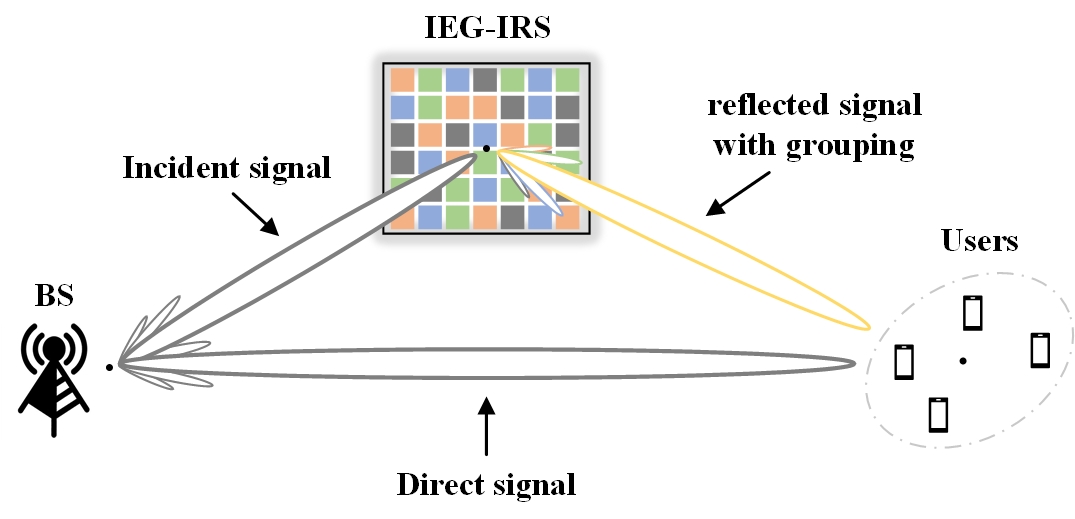}
		\caption{An illustration of the downlink transmission in multi IEG-RISs aided MU-MISO system}
		\label{fig:2}
	\end{figure}
	
	As illustrated in Fig.\ref{fig:2}, we consider an XL-IRS-aided downlink MU-MISO system, where an XL-IRS equipped with $N$ reflecting elements is deployed to assist in the communications from a BS equipped with $M$ antennas to $K$ single-antenna users. To characterize the general performance gains enabled by the proposed IEG-IRS, we adopt the Rician fading channel model for analysis. In this way, an arbitrary channel matrix $\mathbf{H}$ is then given by
	\begin{equation}
		\label{eq:4}
		\mathbf{H} = \delta \big( {\sqrt {\frac{\kappa }{{1 + \kappa }}} \bar{\mathbf{H}} + \sqrt {\frac{1}{{1 + \kappa }}} \tilde{\mathbf{H}} } \big),
	\end{equation}
	where $\delta$ and $\kappa$ are the path loss factor and the Rician factor of $\mathbf{H}$, respectively; ${\bar{\mathbf{H}}}$ denotes the deterministic LoS component, which remains unchanged within the channel coherence time; and ${\tilde{\mathbf{H}}}$ is the stochastic non-line-of-sight (NLoS) component, which each element of that is i.i.d. complex Gaussian random variable with zero mean and unit variance, respectively.
	
	Considering that multi-user linear transmit precoding is employed at BS and each user is assigned with a dedicated beamforming vector. Then, the complex transmit signal at BS can be denoted by $\sum\nolimits_{k = 1}^K {{\mathbf{w}_k}{s_k}}$, where $s_k$ is the symbol transmitted by BS for user $k$ with zero mean and unit variance, and $\mathbf{w}_k \in \mathbb{C}^{M \times 1}$ is the corresponding beamforming vector. Recalling that $\hat{\mathbf{\Phi}} \buildrel\textstyle.\over= \mathrm{diag}\left\{ {\hat{\mathbf{v}}^H \mathbf{G}} \right\} \in \mathbb{C}^{N \times N}$ for simplicity, the dimensions of the combined diagonal reflection-coefficient matrix and the grouping strategy matrix are updated by $\hat{\mathbf{\Phi}} \in \mathbb{R}^{N \times N}$ and $\mathbf{G} \in \mathbb{R}^{Q \times N}$.
	
	Then, the received signal ${y_k} \in \mathbb{C}$ of user $k$ aided by the proposed IEG-IRS can be expressed as
	\begin{equation}
		\label{eq:5}
		{y_k} = \big( {\mathbf{h}_{iu,k}^H \hat{\mathbf{\Phi}} \mathbf{H}_{bi}^H + \mathbf{h}_{bu,k}^H} \big)\sum\limits_{j = 1}^K {{\mathbf{w}_j}{s_j}}  + {z_k},
	\end{equation}
	where $\mathbf{H}_{bi} \in \mathbb{C}^{M \times N}$, $\mathbf{h}_{iu,k} \in \mathbb{C}^{N \times 1}$ and $\mathbf{h}_{bu,k} \in \mathbb{C}^{M \times 1}$ denote the BS-IEG-IRS link, the IEG-IRS-User $k$ link, and the BS-User $k$ link, respectively. $z_k$ denotes additive Gaussian white noise for user $k$ with zero mean and variance $\sigma_k^2$.
	
	It is worth noting that only the reflecting cascade channel and the direct channel can be directly estimated by transmitting limited pilot signals at BS. Then, the cascade channel from the BS to user $k$ via the IRS is usually denoted by ${\mathbf{C}_k} = \mathrm{diag}\big( {\mathbf{h}_{iu,k}^H} \big)\mathbf{H}_{bi}^H \in \mathbb{C}^{N \times M}$. According to \eqref{eq:5}, the received signal $y_k$ of user $k$ can be rewritten by
	\begin{equation}
		\label{eq:6}
		{y_k} = \big( {\hat{\mathbf{v}}^H {\hat{\mathbf{C}}_k} + \mathbf{h}_{bu,k}^H} \big)\sum\limits_{j = 1}^K {{\mathbf{w}_j}{s_j}}  + {z_k},
	\end{equation}
	where ${\hat{\mathbf{C}}_k} = \mathbf{G} \mathbf{C}_k \in \mathbb{C}^{Q \times M}$ denotes the combined cascade channel matrix after the grouping operation at user $k$, and it can be estimated in low pilot overhead conditions. Besides, the optimal grouping strategy $\mathbf{G}$ is predetermined before real-time beamforming.
	
	To illustrate how the IEG-IRS can overcome the two fundamental performance bottlenecks involved in Section I, we will further analyze the asymptotic performance between U-IRS and IEG-IRS in SU-SISO system to reveal the advantages of the proposed grouping strategy in the next section.
	
	\section{Performance Analysis}
	In this section, we analyze the asymptotic performance of an IEG-IRS to reveal its significant performance gains subject to the constraint on the combined dimensions of IRS elements. Further, in order to make the problem analytically tractable and get general and insightful results, we consider an $N$-element IRS aided SU-SISO wireless system with $M=1$ BS antenna and $K=1$ user, while the general MU-MISO case is studied in Section IV.
	
	\subsection{Asymptotic Performance Gain for U-IRS and IEG-IRS}
	To illustrate the performance gains provided by U-IRS and IEG-IRS, the effect of the direct link is ignored by setting $\mathbf{h}_{bu,k}=0$, same as \cite{8811733, 9998527} did. Then, for the generally asymptotic performance of U-IRS and IEG-IRS, we assume Rician-fading channels. Besides, we find that the asymptotic performance gain of the Rician-fading channel for IRS aided SU-SISO system has not yet been proposed. For the above IRS-aided SU-SISO system without a direct link, we simplify the notations by redefining the BS-IRS channel matrix, the IRS-user channel vector, and the BS-IRS-user cascaded channel matrix respectively as
	\begin{equation}
		\label{eq:7}
		\begin{array}{*{20}{l}}
				{{\mathbf{H}_{bi}} := {\mathbf{h}_{bi}} = {\delta _{bi}}\big( {\sqrt {\frac{{{\kappa _{bi}}}}{{1 + {\kappa _{bi}}}}} {{\bar{\mathbf{h}}}_{bi}} + \sqrt {\frac{1}{{1 + {\kappa _{bi}}}}} {{\tilde{\mathbf{h}}}_{bi}}} \big)},\\
				{{\mathbf{h}_{iu,k}} := {\mathbf{h}_{iu}} = {\delta _{iu}}\big( \sqrt {\frac{{{\kappa _{iu}}}}{{1 + {\kappa _{iu}}}}} {{\bar{\mathbf{h}}}_{iu}} + \sqrt {\frac{1}{{1 + {\kappa _{iu}}}}} {{\tilde{\mathbf{h}}_{iu}}} \big)},\\
				{{\mathbf{C}_k} := \mathbf{c} = \mathbf{h}_{iu}^* \odot \mathbf{h}_{bi}^*},
		\end{array}
	\end{equation}
	where $\kappa _{bi}$ and $\kappa _{iu}$, and $\delta_{bi}$ and $\delta_{iu}$ are the Rician factors and the path loss factors of $\mathbf{h}_{bi}$ and $\mathbf{h}_{iu}$, respectively. In particular, the above channel model is reduced to the LoS channel when $\kappa \to \infty$ or Rayleigh fading channel when $\kappa \to 0$. Then, the cascaded channel denoted by $\mathbf{c}$ can be further rewritten as
	\begin{equation}
		\label{eq:8}
		\begin{aligned}
			\mathbf{c} =& \underbrace{\bar{a} {\delta_{bi}}{\delta_{iu}} \bar{\mathbf{h}}_{iu}^* \odot \bar{\mathbf{h}}_{bi}^*}_{{\text{Cascaded Deterministic Component $\mathbf{c}_1$}}}
			\\ &+ \underbrace{\begin{array}{c} \tilde{a}{\delta_{bi}}{\delta_{iu}}\tilde{\mathbf{h}}_{iu}^*\odot\tilde{\mathbf{h}}_{bi}^*
					+ \bar{b}{\delta_{bi}}{\delta_{iu}}\bar{\mathbf{h}}_{iu}^*\odot\tilde{\mathbf{h}}_{bi}^*
					\\+ \tilde{b}{\delta_{bi}}{\delta_{iu}}\tilde{\mathbf{h}}_{iu}^*\odot\bar{\mathbf{h}}_{bi}^* \end{array}}_{{\text{Cascaded Stochastic Component $\mathbf{c}_2$}}},
		\end{aligned}
	\end{equation}
	where  $\bar{a}$, $\tilde{a}$, $\bar{b}$ and $\tilde{b}$ are $\sqrt {\frac{{{\kappa_{bi}\kappa_{iu}}}}{{\left(1 + {\kappa_{bi}}\right)}{\left(1 + {\kappa_{iu}}\right)}}}$, $\sqrt {\frac{{{1}}}{{\left(1 + {\kappa_{bi}}\right)}{\left(1 + {\kappa_{iu}}\right)}}}$, $\sqrt {\frac{{{\kappa_{iu}}}}{{\left(1 + {\kappa_{bi}}\right)}{\left(1 + {\kappa_{iu}}\right)}}}$ and $\sqrt {\frac{{{\kappa_{bi}}}}{{\left(1 + {\kappa_{bi}}\right)}{\left(1 + {\kappa_{iu}}\right)}}}$, respectively, and we let $\mathbf{c}_1$ and $\mathbf{c}_2$ denote the cascaded deterministic and stochastic components, respectively.
	
	The LoS components can be expressed by the responses of the uniform linear array (ULA) and the array response of an $N$-element ULA is
	\begin{equation}
		\label{eq:9}
		{\boldsymbol{\alpha} _N}\big( \theta  \big) = {\big[ 1,e^{j\pi \sin \theta },\cdots,e^{j\left( {N - 1} \right)\pi \sin \theta } \big]^T,}
	\end{equation}
	where $\theta \in \big[ {0,\pi } \big]$ is the angle of departure (AoD) or angle of arrival (AoA) of a transmitting signal. For simplicity, we assume $\sum\nolimits_{n = 0}^{N - 1} {{e^{jn\pi \sin \theta }}}  \to 0$ to obtain essential insight. Under this condition, the LoS components $\bar{\mathbf{h}}_{bi}$ and $\bar{\mathbf{h}}_{iu}$ can be expressed as
	\begin{equation}
		\label{eq:10}
		\left\{ {\begin{array}{*{20}{c}}{{{\bar{\mathbf{h}}}_{bi}} = {\boldsymbol{\alpha} _N}\big( {{\theta _{bi}}} \big)},\\{{{\bar{\mathbf{h}}}_{iu}} = {\boldsymbol{\alpha} _N}\big( {{\theta _{iu}}} \big)},\end{array}} \right.
	\end{equation} 
	where $\theta_{bi}$ and $\theta_{iu}$ are the AoDs from the ULA at BS and IRS, respectively.
	
	For a fair comparison, let $Q$ denote the dimensions of the elements for the two IRSs, which ensures the pilot overhead involved consistent. In order to get insightful results, it is assumed that each group in IEG-IRS contains the same number of elements denoted by $\mu$. Without considering the effect of the transmit power, the normalized transmit power is adopted. Furthermore, given the constraints on the number of pilots, the IEG-IRS is capable of optimizing adjustments for $Q$ groups of elements ($N=Q \times \mu$ IRS elements), whereas U-IRS is restricted to a maximum of $Q$ IRS elements.
	
	Then, for the optimal RCV where the phase alignment solution is given by \cite{8811733}, we have $\left\| {{\check{\mathbf{c}}}} \right\|_1 = \left| {\check{\mathbf{v}}^H{\check{\mathbf{c}}}} \right|$ for the $Q$-element U-IRS and $\left\| {\hat{\mathbf{c}}} \right\|_1 = \left| {\hat{\mathbf{v}}^H\mathbf{G}\mathbf{c}} \right|$ for the $N$-element IEG-IRS, respectively. Therefore, the channel gains achieved by the U-IRS and IEG-IRS are given by
	\begin{subequations}
		\label{eq:11}
		\begin{align}
			&{{\check{\mathcal{G}}} = {{\left\| {{\check{\mathbf{c}}}} \right\|}_1^2},} \label{eq:11A}\\
			&{{\hat{\mathcal{G}}} = {{\left\| {{\hat{\mathbf{c}}}} \right\|}_1^2},} \label{eq:11B}
		\end{align}
	\end{subequations}
	
	Accordingly, the original problem of performance gain maximization for an IEG-IRS aided SU-SISO system can be formulated as
	\begin{equation}
		\label{eq:12}
		\begin{array}{*{20}{l}}
			{{P}:}&{\mathop {\max }\limits_{\mathbf{G}} }{\quad}&{{\hat{\mathcal{G}}},} \\
			{}&{\mathrm{s.t.}}&{\eqref{eq:2A}, \eqref{eq:2B}, \eqref{eq:2C},} 
		\end{array}
	\end{equation}
	Note that problem $P$ is the 0-1 integer programming problem for solving $L_1$ norm, which is challenging to solve. However, on the other hand, it is easy to verify that $\left\| {{\hat{\mathbf{c}}}} \right\|_1 \leq \left\| {{\mathbf{c}}} \right\|_1$ and the optimization of the grouping strategy aims to narrow the performance gap between $\left\| {{\hat{\mathbf{c}}}} \right\|_1$ and $\left\| {{\mathbf{c}}} \right\|_1$. As a consequence, in order to deal with the non-convex $L_1$ norm and 0-1 integer programming in \eqref{eq:12}, we exploit $K$-Nearest Neighbor (KNN) methods proposed in \cite{1053964} to divide the elements of the reflective cascade channel vector into $Q$ groups based on the phase similarity. 
	
	Nevertheless, given the dimensions of XL-IRS elements, it is not feasible to obtain all accurate cascaded CSI in advance. Based on the S-CSI, the reflecting elements will be pre-grouped. Specifically, the proposed grouping strategy is designed to divide the statistical cascaded channel components with close phases into a common group to the greatest extent possible, thereby maximizing the statistical cascaded channel array gains. Consequently, the combined cascaded channel can provide a high channel gain at user under the constraints on the combined dimensions of XL-IRS elements. The idea of our proposed grouping strategy will be verified later.

	Subsequently, the following proposition for a U-IRS aided SU-SISO system, subject to the constraints on the dimensions of IRS elements, is provided.
	
	\textit{Proposition 1 (Asymptotic performance gain for U-IRS)}: Assuming ${\mathbf{h}_{bi}}  \sim {\mathcal C}{\mathcal N} \big( \delta_{bi} \sqrt {\frac{\kappa_{bi}}{1 + {\kappa_{bi}}}} {\boldsymbol{\alpha} _Q}\left( \theta_{bi} \right), \frac{\delta_{bi}^2}{1 + \kappa_{bi}} {\mathbf{I}_Q} \big)$, ${\mathbf{h}_{iu}} \sim {\mathcal C}{\mathcal N}\big( {{\delta_{iu} \sqrt {\frac{{{\kappa_{iu}}}}{{1 + {\kappa_{iu}}}}} {\boldsymbol{\alpha} _Q}\left( {{\theta_{iu}}} \right)}, \frac{{{\delta_{iu}^2}}}{{1 + {\kappa_{iu}}}} {\mathbf{I}_Q}} \big)$, the cascaded channel $\check{\mathbf{c}}$ perfectly known and letting $Q \to \infty $, the asymptotic channel gain $\check{\mathcal{G}}$ for a U-IRS aided SU-SISO system is given by,
	\begin{equation}
		\begin{split}
			\label{eq:13}
			{\check{{\mathcal{G}}}} \to {Q^2}\frac{{{\pi ^2}\delta_{bi}^2\delta_{iu}^2\tilde{a}^2}}{{16}}{L_{1/2}^2\big( { - {\kappa_{bi}}} \big)L_{1/2}^2\big( { - {\kappa_{iu}}} \big)},
		\end{split}
	\end{equation}
	where ${L_{\frac{1}{2}}}\left( -x \right) = {e^{-\frac{x}{2}}}\big[ {\big( {1 + x} \big){I_0}\big( {\frac{x}{2}} \big) + x{I_1}\big( {\frac{x}{2}} \big)} \big]$, $I_0\left(\cdot\right)$ and $I_1\left(\cdot\right)$ denote exponentially scaled modified Bessel function of orders 0 and 1, respectively.
	\begin{IEEEproof}
		For optimal $\check{\mathbf{v}}$, we have $\left\| {{\check{\mathbf{c}}}} \right\|_1 = \left\| {\mathbf{h}_{iu}^* \odot \mathbf{h}_{bi}^*} \right\|_1 = \sum\nolimits_{q = 1}^Q {\left| {{h_{1,q}}} \right|} \left| {{h_{2,q}}} \right|$, where $h_{1,q}$ and $h_{2,q}$ are the $q$-th elements in $\mathbf{h}_{bi}$ and $\mathbf{h}_{iu}$, respectively. Since ${\left| {h_{1,q}} \right|}$ and ${\left| {h_{2,q}} \right|}$ are statistically independent and follow Rician distribution with mean values $\frac{{\sqrt \pi  }}{2}{\delta_{bi}}\sqrt {\frac{1}{{1 + {\kappa_{bi}}}}} {L_{1/2}}\big( { - {\kappa_{bi}}} \big)$ and $\frac{{\sqrt \pi  }}{2}{\delta_{iu}}\sqrt {\frac{1}{{1 + {\kappa_{iu}}}}} {L_{1/2}}\big( { - {\kappa_{iu}}} \big)$, respectively, we have $\mathbb{E}\big({\left| {{h_{1,q}}} \right|} \left| {{h_{2,q}}} \right|\big) = \frac{{\pi  }}{4}{\delta_{bi}}{\delta_{iu}}\tilde{a} {L_{1/2}}\big( { - {\kappa_{bi}}} \big){L_{1/2}}\big( { - {\kappa_{iu}}} \big)$. By using the fact that $\sum\nolimits_{q = 1}^Q {\left| {{h_{1,q}}} \right|} \left| {{h_{2,q}}} \right|/Q \to \frac{{\pi  }}{4}{\delta_{bi}}{\delta_{iu}}\tilde{a} {L_{1/2}}\big( { - {\kappa_{bi}}} \big){L_{1/2}}\big( { - {\kappa_{iu}}} \big)$ as $Q \to \infty$ with \eqref{eq:11A}, it follows that ${\check{\mathcal{G}}} \to {Q^2}\frac{{{\pi ^2}\delta_{bi}^2\delta_{iu}^2\tilde{a}^2}}{{16}}{L_{1/2}^2\big( { - {\kappa_{bi}}} \big)L_{1/2}^2\big( { - {\kappa_{iu}}} \big)}$. This completes the proof.
	\end{IEEEproof}
	
	From \eqref{eq:13}, by using the fact that ${L_{\frac{1}{2}} }\big( 0 \big) = 1$ and $\mathop {\lim }\limits_{x \to  - \infty } {L_\nu }\big( x \big) = \frac{{{{\left| x \right|}^\nu }}}{{\Gamma \left( {1 + \nu } \right)}}$ where $\Gamma \left(  \cdot  \right)$ denotes Gamma function and $\Gamma \big( {\frac{3}{2}} \big) = \frac{{\sqrt \pi  }}{2}$, we have $\mathop {\lim }\limits_{\kappa \to 0} {\check{\mathcal{G}}} \to {Q^2}\frac{{{\pi ^2}\delta_{bi}^2\delta_{iu}^2}}{{16}}$ and $\mathop {\lim }\limits_{\kappa \to \infty } {\check{\mathcal{G}}} \to Q^2\delta_{bi}^2\delta_{iu}^2 $. Therefore, regardless of whether the link is purely LoS, NLoS, or a combination of both, the asymptotic channel gain of the U-IRS aided SU-SISO system exhibits an appealing ``squared gain''. However, as the dimension $Q$ of the cascaded channel increases, the U-IRS cannot circumvent those critical performance constraints, including the pilot overhead in channel estimation and the high dimensional matrix operations required for real-time precoding.
	
	Observing \eqref{eq:8}, it is not difficult to verify that the cascaded deterministic and stochastic components are statistically independent. The IRS elements can be pre-grouped based on the phase similarity of the cascaded deterministic component. Subsequently, the following lemma presents the distribution of the combined cascaded channel.
	
	\textit{Lemma 1 (Distribution for combined cascaded channel)}: Assuming ${\mathbf{h}_{bi}}  \sim {\mathcal C}{\mathcal N}\big( {{\delta_{bi} \sqrt {\frac{{{\kappa_{bi}}}}{{1 + {\kappa_{bi}}}}} {\boldsymbol{\alpha} _N}\big( {{\theta_{bi}}} \big)}, \frac{{{\delta_{bi}^2}}}{{1 + {\kappa_{bi}}}} {\mathbf{I}_N}} \big)$, ${\mathbf{h}_{iu}} \sim {\mathcal C}{\mathcal N}\big( {{\delta_{iu} \sqrt {\frac{{{\kappa_{iu}}}}{{1 + {\kappa_{iu}}}}} {\boldsymbol{\alpha} _N}\big( {{\theta_{iu}}} \big)}, \frac{{{\delta_{iu}^2}}}{{1 + {\kappa_{iu}}}} {\mathbf{I}_N}} \big)$ and letting $\mu \to \infty$, the distribution for the combined cascaded channel $\hat{\mathbf{c}}$ is given by
	\begin{equation}
		\label{eq:14}
		\hat{\mathbf{c}} \sim \mathcal{CN}\big( \mu{\delta_{bi}}{\delta_{iu}}\imath \bar{a}\boldsymbol{\vartheta} ,\mu \delta_{bi}^2\delta_{iu}^2\big( {1 - \bar{a}^2}\big){\mathbf{I}_Q} \big),
	\end{equation}
	where $\boldsymbol{\vartheta} = {\big[e^{-j\frac{{\pi }}{Q}},\cdots ,e^{-j\frac{{\left( {2Q - 1} \right)\pi }}{Q}}\big]^H }$ and $\imath = {\frac{{\sin \frac{\pi }{Q}}}{{\frac{\pi }{Q}}}}$.
	\begin{IEEEproof}
		Please see Appendix A.
	\end{IEEEproof}
	
	Therefore, by using the proposed grouping strategy, the combined cascaded channel $\hat{\mathbf{c}}$ can be regarded as a new Rician channel. For comparison, under the same transmission conditions, the asymptotic channel gain of an IEG-IRS-aided SU-SISO system with the constraint of the combined dimensions of XL-IRS elements is proposed in the following proposition.
	
	\textit{Proposition 2 (Asymptotic performance gain for IEG-IRS)}: Assuming ${\mathbf{h}_{bi}}  \sim {\mathcal C}{\mathcal N}\big( {{\delta_{bi} \sqrt {\frac{{{\kappa_{bi}}}}{{1 + {\kappa_{bi}}}}} {\boldsymbol{\alpha} _N}\big( {{\theta_{bi}}} \big)}, \frac{{{\delta_{bi}^2}}}{{1 + {\kappa_{bi}}}} {\mathbf{I}_N}} \big)$, ${\mathbf{h}_{iu}} \sim {\mathcal C}{\mathcal N}\big( {{\delta_{iu} \sqrt {\frac{{{\kappa_{iu}}}}{{1 + {\kappa_{iu}}}}} {\boldsymbol{\alpha} _N}\big( {{\theta_{iu}}} \big)}, \frac{{{\delta_{iu}^2}}}{{1 + {\kappa_{iu}}}} {\mathbf{I}_N}} \big)$, the combined cascaded channel $\hat{\mathbf{c}}$ in \eqref{eq:14} perfectly known, and letting $Q \to \infty $ and $\mu \to \infty$, the asymptotic channel gain ${\hat{\mathcal{G}}}$ for an IEG-IRS aided SU-SISO system is given by,
	\begin{equation}
		\begin{split}
			\label{eq:15}
			{\hat{\mathcal{G}}} \to NQ \frac{{\pi \delta_{bi}^2\delta_{iu}^2}}{4}\big(1 - \bar{a}^2\big) L_{1/2}^2\big( { - \frac{\bar{a}^2}{{1 - \bar{a}^2}}\mu } \big),
		\end{split}
	\end{equation}
	\begin{IEEEproof}
		 For optimal $\hat{\mathbf{v}}$, we have $\left\| {{\hat{\mathbf{c}}}} \right\|_1 = \sum\nolimits_{q = 1}^Q {\left| {{\hat{c}_{q}}} \right|}$,  where $\hat{c}_{q}$ is the $q$-th elements in $\hat{\mathbf{c}}$. Since ${\left| {{{\hat c}_q}} \right|}$ follows Rician distribution with mean values $\frac{{\sqrt \pi  }}{2}{\delta_{bi}}{\delta_{iu}}\sqrt {\big(1 - \bar{a}^2\big) \mu} {L_{1/2}}\big( { - \frac{\imath^2\bar{a}^2}{{1 - \bar{a}^2}}\mu } \big)$, we have ${\sum\nolimits_{q = 1}^Q {\left| {{{\hat c}_q}} \right|} }/Q \to \frac{{\sqrt \pi  }}{2}{\delta_{bi}}{\delta_{iu}}\sqrt {\big(1 - \bar{a}^2\big) \mu} {L_{1/2}}\big( { - \frac{\bar{a}^2}{{1 - \bar{a}^2}}\mu } \big)$ as $Q \to \infty$. By using this fact with (10b), it follows that ${\hat{\mathcal{G}}} \to NQ \frac{{\pi \delta_{bi}^2\delta_{iu}^2}}{4}\big(1 - \bar{a}^2\big) L_{1/2}^2\big( { - \frac{\bar{a}^2}{{1 - \bar{a}^2}}\mu } \big)$. This completes the proof.
	\end{IEEEproof}
	
	From \eqref{eq:15}, given the number of group $Q$, the significant performance gain achieved by the proposed IEG-IRS can be obtained by increasing the number of reflecting elements $\mu$ for each group. Further, by using the fact that $\mathop {\lim }\limits_{{{\bar{a}^2}\mu } \to \infty} L_{1/2}^2\big( { - \frac{\bar{a}^2}{{1 - \bar{a}^2}}\mu } \big) \to \frac{4\mu}{\pi}\frac{\bar{a}^2}{{1 - \bar{a}^2}}$, it follows that
	\begin{equation}
		\begin{split}
			\label{eq:16}
			{\hat{\mathcal{G}}} \to \begin{matrix}
				{N^2}\delta_{bi}^2\delta_{iu}^2\bar{a}^2, &\mathrm{as}\ {{\bar{a}^2}\mu } \to \infty,
			\end{matrix}
		\end{split}
	\end{equation}
	Note that only when $\mu \to \infty$ and $\bar{a} \nrightarrow 0$, it satisfies $\bar{a}^2\mu \to \infty$.
	
	Furthermore, when the number of groups is small, Proposition 2 is no longer applicable. According to \eqref{eq:14}, it is not difficult to verify that the Rician factor of the combined cascaded channel $\hat{\mathbf{c}}$ satisfies $\kappa = \frac{\bar{a}^2\imath^2}{\left( {1 - \bar{a}^2}\right)}\mu$. As such, when $\bar{a} \nrightarrow 0$, and letting $\mu \to \infty$ and $Q > 1$, we have $\kappa \to \infty$ and the combined cascaded channel can be considered as a ``deterministic virtual LoS'' channel approximatively, as
	\begin{equation}
		\label{eq:17}
		\hat{\mathbf{c}} \to \begin{matrix}
			\mu{\delta_{bi}}{\delta_{iu}}\imath\bar{a}\boldsymbol{\vartheta}, &\mathrm{as}\ {{\bar{a}^2}\mu } \to \infty \  \text{and} \  Q > 1,
		\end{matrix}
	\end{equation}
	
	Then, for $Q = 1$, we have $\hat{\mathbf{c}} \sim \mathcal{CN}\big(0 , \delta_{bi}^2\delta_{iu}^2\big( {1 - \bar{a}^2}\big){\mathbf{I}_N} \big)$ as $N \to \infty$ from \eqref{eq:14}. Thus, by using the above fact and \eqref{eq:17}, when $\bar{a} \nrightarrow 0$ and $\mu \to \infty$, the asymptotic performance gain of an IEG-IRS aided SU-SISO system without the constraint that $Q \to \infty$ is presented by:
	\begin{equation}
		\begin{split}
			\label{eq:18}
			{\hat{\mathcal{G}}} \to \begin{matrix}
				\left\{\begin{matrix}
					{N\delta_{bi}^2\delta_{iu}^2\big(1 - \bar{a}^2\big)},&{Q = 1},\\
					{{N^2}\delta_{bi}^2\delta_{iu}^2{{\imath}^2}\bar{a}^2},&{Q > 1},
				\end{matrix}\right.& \mathrm{as}\  {{\bar{a}^2}\mu } \to \infty,
			\end{matrix}
		\end{split}
	\end{equation}

	From \eqref{eq:14} and \eqref{eq:18}, it is not difficult to observe that when $Q > 1$, the asymptotic channel gain is proportional to $N^2$ and largely brought by the deterministic LoS component of the combined cascaded channel $\hat{\mathbf{c}}$. As ${{\bar{a}^2}\mu } \to \infty$, the optimal combined RCV is obtained by $\hat{\mathbf{v}} = \mathrm{arg}\big(\boldsymbol{\vartheta}\big)$ without real-time optimization. Then, as $Q$ ranges from 2 to 4, the performance gap compared with \eqref{eq:16} decreases from $59\%$ to $19\%$ rapidly. As $Q \to \infty$, we attain the same result as \eqref{eq:16}. Besides, for a monotonic characteristic of logical function, as the total number of elements is sufficiently large, the system capacity gap caused by a small number of groups can be ignored. It indicates that when a LoS component is available, the IEG-IRS can still deliver significant performance gain (similar to the ``squared gain'' property in \cite{8811733}) in low pilot overhead conditions. 
	
	In the next subsection, a performance loss analysis of the asymptotic channel gains achieved by IEG-IRS is proposed to reveal the superiority of the IEG-IRS in the SU-SISO system.
	
	\subsection{Performance Loss for IEG-IRS}
	Assuming that the dimensions of U-IRS elements are no longer constrained, the asymptotic channel gain achieved by U-IRS can be regarded as the upper bound of the performance gain achieved by IEG-IRS according to \eqref{eq:13}. Then, the following proposition presents the analysis of the performance loss for IEG-IRS.
	
	\textit{Proposition 3 (Performance loss for IEG-IRS)}: Assuming ${\mathbf{h}_{bi}}  \sim {\mathcal C}{\mathcal N}\big( {{\delta_{bi} \sqrt {\frac{{{\kappa_{bi}}}}{{1 + {\kappa_{bi}}}}} {\boldsymbol{\alpha} _N}\big( {{\theta_{bi}}} \big)}, \frac{{{\delta_{bi}^2}}}{{1 + {\kappa_{bi}}}} {\mathbf{I}_N}} \big)$, ${\mathbf{h}_{iu}} \sim {\mathcal C}{\mathcal N}\big( {{\delta_{iu} \sqrt {\frac{{{\kappa_{iu}}}}{{1 + {\kappa_{iu}}}}} {\boldsymbol{\alpha} _N}\big( {{\theta_{iu}}} \big)}, \frac{{{\delta_{iu}^2}}}{{1 + {\kappa_{iu}}}} {\mathbf{I}_N}} \big)$, the combined cascaded channel $\hat{\mathbf{c}}$ perfectly known, and letting $Q \to \infty$ and $\mu \to \infty $, the performance loss $L$ of the asymptotic channel gain for IEG-IRS is given by
	\begin{equation}
		\begin{split}
			\label{eq:19}
			L = 1 - \frac{4{\big( {1 + {\kappa_{bi}} + {\kappa_{iu}}} \big)}L_{1/2}^2\big( { - \frac{\kappa_{bi}\kappa_{iu}}{{1 + \kappa_{bi} + \kappa_{iu}}}\mu } \big)}{\pi{L_{1/2}^2\big( { - {\kappa_{bi}}} \big)L_{1/2}^2\big( { - {\kappa_{iu}}} \big)}\mu},
		\end{split}
	\end{equation}
	\begin{IEEEproof}
		By combining Proposition 1 and Proposition 2, we have the ratio of performance gain IEG-IRS to U-IRS denoted by $\frac{4{\big( {1 + {\kappa_{bi}} + {\kappa_{iu}}} \big)}L_{1/2}^2\big( { - \frac{\kappa_{bi}\kappa_{iu}}{{1 + \kappa_{bi} + \kappa_{iu}}}\mu } \big)}{\pi{L_{1/2}^2\big( { - {\kappa_{bi}}} \big)L_{1/2}^2\big( { - {\kappa_{iu}}} \big)}\mu}$. This thus completes the proof.
	\end{IEEEproof}
	
	From \eqref{eq:19}, we observe that when the Rician factor and the number of groups are sufficiently large, the performance loss for IEG-IRS is negligible. Furthermore, by properly designing the deployment, the IRS is capable of providing additional virtual LoS links with the user in the NLoS areas \cite{9737357}. In this context, the Rician factors of both the BS-IRS link and IRS-user link are generally large. Therefore, in the ideal case ($\kappa \to \infty$ and $Q \to \infty$), there is no performance loss in the asymptotic channel gain for IEG-IRS.
	
	\section{General Multiuser System}
	
	To investigate the capacity gain achieved by the IEG-IRS in typical communication scenarios, in this section, we consider a general XL-IRS aided MU-MISO system. Specifically, we formulate the problem of WSR maximization in Subsection IV-A. Then, in Subsection IV-B, a two-stage algorithm for optimizing the beam-domain grouping strategy and the multi-user active-passive beamforming is proposed to solve the formulated problem.
	
	\subsection{WSR Maximization Problem Formulation}
	According to the IEG-IRS aided MU-MISO system described in Subsection II-C, the signal-to-interference-plus-noise ratio (SINR) at user $k$ can be represented by
	\begin{equation}
		\label{eq:20}
		{\gamma _k} = \frac{{{{\left| {{{\mathbf{h}}_k^H}{\mathbf{w}_k}} \right|}^2}}}{{\sum\nolimits_{j \ne k}^K {{{\left| {{{\mathbf{h}}_k^H}{\mathbf{w}_j}} \right|}^2} + {\sigma ^2}} }},
	\end{equation}
	where ${\mathbf{h}}^H_k = {\hat{\mathbf{v}}^H {\mathbf{G}\mathbf{C}_k} + \mathbf{h}_{bu,k}^H} \in \mathbb{C}^{1 \times M}$ denotes the superimposed channel from both the BS-user $k$ (direct) link and the BS-IEG-IRS-user $k$ (reflect) link.
	
	It is worth emphasizing that the use of XL-IRS makes both the channel estimation and real-time beamforming inoperable at BS. Inspired by performance analysis presented in Section III, we propose a beam-domain grouping method and the optimal grouping strategy $\mathbf{G}$ is predetermined by the multi-user S-CSI. Then the combined cascaded channel available for user $k$ denoted by $\hat{\mathbf{C}}_k = \mathbf{G}\mathbf{C}_k \in \mathbb{C}^{Q \times M}$ can be estimated in real time. In contrast to the SU-SISO case, the deterministic component of the BS-user link has a direct impact on the accuracy of the grouping strategy optimization, while the optimization of the combined RCV is affected by its stochastic component. From \eqref{eq:15} and \eqref{eq:18}, the IEG-IRS is capable of maintaining notable capacity gains while simultaneously reducing high signaling overhead and computational complexity introduced by the XL-IRS. Specifically, in low pilot overhead conditions, the array gains of each group and the system capacity can keep improving as the total number of IRS elements increases. Therefore, utilizing those above characteristics, we investigate an IEG-IRS aided MU-MISO system.
	
	Then, our objective is to maximize the WSR of the IEG-IRS aided MU-MISO system by jointly optimizing the grouping strategy $\mathbf{G}$, the combined dimensions of IRS elements $Q$, the transmit beamforming $\mathbf{w}$ and the combined RCV $\hat{\mathbf{v}}$. Therefore, the original problem is formulated as
	\begin{subequations}
		\label{eq:21}
		\begin{align}
			{{P_0}:}{\quad}{\mathop {\max }\limits_{\mathbf{G},Q,\mathbf{w},\hat{\mathbf{v}}}}{\quad}&{\sum\limits_{k = 1}^K {\varpi_k{{\log }_2}\big( {1 + {\gamma _k}} \big)} }, \label{eq:21A}\\
			{\mathrm{s.t.}}{\quad}{\quad}&{{\eqref{eq:2A}},{\eqref{eq:2B}}, {\eqref{eq:2C}}}, \nonumber\\
			&{Q \le {Q_0} \ll N}, \label{eq:21B}\\
			&{{{\left\| \mathbf{w} \right\|}^2} \le P_{BS}^{\max }}, \label{eq:21C}\\
			&{\left| {{{\hat v}_q}} \right| = 1}, \label{eq:21D}
		\end{align}
	\end{subequations}
	where $\varpi_k > 0$, $\mathbf{w} = {\big[ {{{\mathbf{w}_1^T}},\cdots, {{\mathbf{w}_K^T}}} \big]^T} \in \mathbb{C}^{MK \times 1}$ and $Q_0$ denote the weight for user $k$, the overall transmit beamforming vector for $K$ users and the maximum combined dimensions of IRS elements supported by the system, respectively; constrains \eqref{eq:21C} and \eqref{eq:21D} denote the power constraint at the BS and the unit-modulus constraint of the combined IRS reflection coefficients, respectively.
	
	It can be observed that problem $P_0$ is challenging to solve due to both the non-convexity and highly coupled variables. In particular, the introduction of the IEG-IRS brings the 0-1 integer grouping matrix $G$, which is difficult to optimize and presents the active-passive beamforming vectors from updating in real time. Therefore, to efficiently solve this problem, we propose a two-stage algorithm to optimize the beam-domain grouping strategy and the multi-user active-passive beamforming based on the alternating optimization, quadratic transform for FP and majorization-minimization algorithm. Besides, inspired by the idea of the proposed grouping strategy in an IEG-IRS aided SU-SISO system, we propose a straightforward and effective heuristic grouping method. Those details will be provided in the next subsection.

	\subsection{Proposed Two-Stage Algorithm Design}
	To deal with the high coupling variables and the non-convex sum-of-logarithms with fractions in the objective function of problem $P_0$, in this section, we exploit quadratic transform for FP method proposed in \cite{8314727} to decouple the variables in problem $P_0$, then enable the separate optimization of multiple variables. Since the channel gain is positively correlated with the number of groups, we have $Q=Q_0$. Thus, the original problem $P_0$ can be reformulated as
	\begin{equation}
		\label{eq:22}
		\begin{array}{*{20}{l}}
			{{P_1}:}&{\mathop {\max }\limits_{\mathbf{G},\mathbf{w},\hat{\mathbf{v}},\boldsymbol{\varsigma} ,\boldsymbol{\xi} } }&{\sum\limits_{k = 1}^K {\varpi_k{{\log }_2}\big( {1 + {{\varsigma} _k}} \big) - \varpi_k{{\varsigma} _k}} } \\
			{}&{}&{\quad\ \ }{ + 2\sqrt {\varpi_k\big( {1 + {{\varsigma} _k}} \big)} \Re \left\{ {{\xi} _k^*\mathbf{h}_k^H{\mathbf{w}_k}} \right\}} \\
			{}&{}&{\quad\ \ }{ - {{\left| {{{\xi} _k}} \right|}^2}\big( {\sum\limits_j^K {{{\left| {\mathbf{h}_k^H{\mathbf{w}_j}} \right|}^2}}  + {\sigma ^2}} \big)}, \\
			{}&{\quad\mathrm{s.t.}}&{{\eqref{eq:2A}}, {\eqref{eq:2B}}, {\eqref{eq:2C}}, {\eqref{eq:21C}}, {\eqref{eq:21D}}},
		\end{array}
	\end{equation}
	where $\boldsymbol{\varsigma}  = \big[ {{{\varsigma _1}},\cdots, {{\varsigma _K}}} \big] \in \mathbb{R}_ + ^K$ and $\boldsymbol{\xi}  = \big[ {{{\xi _1}},\cdots,{{\xi _K}}} \big] \in {\mathbb{C}^K}$ denote the auxiliary variables, respectively.
	
	It can be observed that the updates in each iteration step of the grouping strategy matrix $\mathbf{G}$, beamforming vector $\mathbf{w}$, combined RCV $\hat{\mathbf{v}}$, auxiliary variables $\boldsymbol{\varsigma}$ and $\boldsymbol{\xi}$ in problem $P_1$ are all optimal. Then, a locally optimal solution to \eqref{eq:22} can be obtained by alternately optimizing these variables until convergence.
	
	However, the dimension of cascaded channel $\mathbf{C}_k$ for user $k$ is prohibitively large. Inspired by the combined channel gain maximization problem \eqref{eq:12} in Section III, we propose a two-stage algorithm by separating problem $P_1$ into two parts, for optimizing the grouping strategy and active-passive precoding, respectively. Specifically, the grouping strategy at the IEG-IRS is optimized in the first stage by utilizing the multi-user S-CSI. Then, in the second stage, we propose a joint transmit beamforming and combination-reflection precoding scheme to solve the WSR maximization problem with given grouping strategy matrix $\mathbf{G}$ under the combined multi-user I-CSI.
	
	\subsubsection{Beam-Domain Grouping Method}
	
	Motivated by the grouping idea employed in an IEG-IRS aided SU-SISO system, we investigate the grouping strategy in MU-MISO system through the multi-user S-CSI. Let $\bar{\mathbf{C}}_k \in \mathbb{C}^{N \times M}$ and $\bar{\mathbf{h}}_{bu,k} \in \mathbb{C}^{M \times 1}$ denote the S-CSI of the cascade channel and direct channel for $k$ user, respectively. Assuming that the optimal statistical active-passive precoding vectors are known in advance, then we focus on optimizing the grouping strategy matrix $\mathbf{G}$. Let $\bar{\mathbf{w}} = {\big[ \bar{\mathbf{w}}_1^T,\cdots,\bar{\mathbf{w}}_K^T \big]^T} \in \mathbb{C}^{MK \times 1}$ and $\bar{\mathbf{v}} = {\big[ e^{j{\bar{\theta} _1}},\cdots,e^{j{\bar{\theta} _Q}} \big]^T} \in \mathbb{C}^{Q \times 1}$ denote the optimal transmit beamforming vector and combined RCV for the statistical channel model. Recalling problem $P_1$, the optimization problem of the grouping strategy matrix can be reformulated as
	\begin{equation}
		\label{eq:23}
		\begin{array}{*{20}{l}}
			{{P_2}:}&{\mathop {\max }\limits_{\mathbf{G}} }&{\sum\limits_{k = 1}^K {2\sqrt {\varpi_k\big( {1 + {\varsigma _k}} \big)} \Re \left\{ {\xi _k^*{{{\bar{\mathbf{v}}}}^H}{\mathbf{G}}{\bar{\mathbf{C}}_k}{{\bar{\mathbf{w}}}_k}} \right\}} } \\
			{}&{}&{\quad\ \ }{ - {{\left| {{\xi _k}} \right|}^2}\sum\limits_j^K {{{\left| {\big( {{{{\bar{\mathbf{v}}}}^H}{\mathbf{G}}{\bar{\mathbf{C}}_k} + \bar{\mathbf{h}}_{bu,k}^H} \big){{\bar{\mathbf{w}}}_j}} \right|}^2}} }, \\
			{}&{\mathrm{s.t.}}&{{\eqref{eq:2A}},{\eqref{eq:2B}}, {\eqref{eq:2C}},}
		\end{array}
	\end{equation}
	
	It observes from \eqref{eq:23} that problem $P_1$ is the 0-1 integer programming, which is typically challenging to obtain the optimal solution. Besides, different from an IEG-IRS aided SU-SISO system, the grouping strategy and multi-user transmit beamforming are coupled in which the optimization of the transmit precoding will result in alterations to the beam-domain S-CSI. Thus, to tackle the above issues, we exploit the efficient $L_{21}$-Norm maximization method proposed in \cite{9435935} and alternating optimization algorithm to relax the non-convex constraint represented in \eqref{eq:2A}, so that the grouping strategy matrix $\mathbf{G}$ can be optimized. Specifically, by introducing variable $\boldsymbol{\Gamma} \in \mathbb{R}^{Q \times N}$ where ${\boldsymbol{\Gamma} _n} = \frac{{{\mathbf{G}_n}}}{{{{\left\| {{\mathbf{G}_n}} \right\|}}}}$, and ${\mathbf{G}_n}$ denotes the $n$-th column vector of $\mathbf{G}$, the optimization problem of the grouping strategy matrix can be reformulated as
	\begin{equation}
		\label{eq:24}
		\begin{array}{*{20}{l}}
			{{P_{2.1}}:}&{\mathop {\max }\limits_\mathbf{G} }&{\sum\limits_{k = 1}^K {2\sqrt {\varpi_k\big( {1 + {\varsigma _k}} \big)} \Re \left\{ {\xi _k^*{{{\bar{\mathbf{v}}}}^H}{\mathbf{G}}{\bar{\mathbf{C}}_k}{{\bar{\mathbf{w}}}_k}} \right\}} }\\
			{}&{}&{\quad\ \ }{ - {{\left| {{\xi _k}} \right|}^2}\sum\limits_j^K {{{\left| {\big( {{{{\bar{\mathbf{v}}}}^H}{\mathbf{G}}{\bar{\mathbf{C}}_k} + \bar{\mathbf{h}}_{bu,k}^H} \big){{\bar{\mathbf{w}}}_j}} \right|}^2}} }\\
			{}&{}&{ + \sum\limits_{n = 1}^N {{\boldsymbol{\Gamma} _n^T}{\mathbf{G}_n}} },\\
			{}&{\mathrm{s.t.}}&{{\eqref{eq:2B}},\eqref{eq:2C},}
		\end{array}
	\end{equation}
	As problem $P_{2.1}$ is a convex quadratic program (QP), it can be optimally solved by existing convex optimization solvers such as CVX \cite{cvx}. In addition, similar to the single-user case, we also propose a heuristic reflection-coefficient approximation method for an IEG-IRS aided MU-MISO system, which narrows the gap between the combined RCV and the optimal statistical RCV.
	
	\subsubsection{Joint Transmit Beamforming and Combination-Reflection Precoding Scheme}
	In this stage, the combined cascaded channel for user $k$ denoted by $\hat{\mathbf{C}}_k  \in \mathbb{C}^{Q \times M}$ can be estimated in real time. Then, recalling ${\mathbf{h}}^H_k = {\hat{\mathbf{v}}^H {\hat{\mathbf{C}}_k} + \mathbf{h}_{bu,k}^H}$, problem $P_1$ is converted into the following three subproblems for given grouping strategy matrix $\mathbf{G}$ separately.
	\paragraph{Fix $\left( {\mathbf{w},\hat{\mathbf{v}}} \right)$ and then optimize $\boldsymbol{\xi}$ and $\boldsymbol{\varsigma}$}
	
	First, after fixing $\mathbf{w},\hat{\mathbf{v}}$ and $\boldsymbol{\varsigma}$, problem $P_1$ can be converted as
	\begin{equation}
		\label{eq:25}
		\begin{array}{*{20}{l}}
			{{P_{3.1}}:} &{\mathop {\max}\limits_{\boldsymbol{\xi}}} &{\sum\limits_{k = 1}^K {2\sqrt{\varpi_k\big( {1 + {\varsigma _k}} \big)} \Re \left\{ {\xi _k^*{\mathbf{h}}_k^H{{\mathbf{w}}_k}} \right\}} }\\
			{}&{}&{\quad\ \ }{ - {{\left| {{\xi _k}} \right|}^2}\big( {\sum\nolimits_{j = 1}^K {{{\left| {{\mathbf{h}}_k^H{{\mathbf{w}}_j}} \right|}^2}}  + {\sigma ^2}} \big)},
		\end{array}
	\end{equation}
	Since problem $P_{3.1}$ in \eqref{eq:25} is a convex unconstrained QP problem, the optimal $\boldsymbol{\xi}$ can be obtained by taking the derivative of the objective function and setting it zero, as
	\begin{equation}
		\label{eq:26}
		{\xi _k} = \frac{{\sqrt {\varpi_k\big( {1 + {\varsigma _k}} \big)} {\omega_k}}}{{{\chi_k}}},
	\end{equation}
	where ${\omega_k} = \mathbf{h}_k^H{\mathbf{w}_k}$ and ${\chi_k} = \sum\nolimits_{j = 1}^K {{{\left| {\mathbf{h}_k^H{\mathbf{w}_j}} \right|}^2} + {\sigma ^2}} $
	
	Then, with $\boldsymbol{\xi}$, $\mathbf{w}$, and $\mathbf{\hat{v}}$ fixed, problem $P_1$ can be converted to:
	\begin{equation}
		\label{eq:27}
		\begin{array}{*{20}{l}}
			{{P_{3.2}}:}&{\mathop {\max }\limits_{\boldsymbol{\varsigma}}  }&{\sum\limits_{k = 1}^K {{\varpi_k{\log }_2}\big( {1 + {\varsigma _k}} \big) - \varpi_k{\varsigma _k}} }\\
			{}&{}&{\quad\ \ }{ + 2\sqrt {\varpi_k\big( {1 + {\varsigma _k}} \big)} \Re \left\{ {\xi _k^*\omega_k} \right\}},
		\end{array}
	\end{equation}
	Since problem $P_{3.2}$ in \eqref{eq:27} is a convex unconstrained programming problem, by adopting the derivative method, the optimal $\boldsymbol{\varsigma}$ can be obtained as
	\begin{equation}
		\label{eq:28}
		\varsigma _k = \frac{{\varepsilon _k^2 + {\varepsilon _k}\sqrt {\varepsilon _k^2 + 4} }}{2},
	\end{equation}
	where $\varepsilon _k=\Re \left\{\frac{{\xi_k^{*}}\omega_k}{\sqrt{\varpi_k}} \right\}$.
	
	We further analysis and decouple the solutions of the auxiliary variables  $\boldsymbol{\xi}$ and $\boldsymbol{\varsigma}$. By substituting \eqref{eq:26} into \eqref{eq:28}, the optimal solutions containing CSI and beamforming vector are given by
	\begin{subequations}
		\label{eq:29}
		\begin{align}
			&{{\xi _k^{opt}} = {A_k}{e^{j{\theta _k}}},} \label{eq:29A}\\
			&{\varsigma _k^{opt} = \frac{{B_k^2 + {B_k}\sqrt {B_k^2 + 4} }}{2},} \label{eq:29B}
		\end{align}
	\end{subequations}
	where ${A_k} = \frac{{\left| {{\omega_k}} \right|}}{{\sqrt {\chi_k^2 - {{\left| {{\omega_k}} \right|}^2}{\chi_k}} }}$, ${{B_k} = \frac{{{{\left| {{\omega_k}} \right|}^2}}}{{\sqrt {\chi_k^2 - {{\left| {{\omega_k}} \right|}^2}{\chi_k}} }}}$ and ${\theta _k} = \arg \big( {\omega_k} \big)$, respectively.
	
	\paragraph{Fix $\left( {\hat{\mathbf{v}}, \boldsymbol{\varsigma}, \boldsymbol{\xi}} \right)$ and then optimize $\mathbf{w}$}
	
	After fixing $\hat{\mathbf{v}}, \boldsymbol{\varsigma}$ and $\boldsymbol{\xi}$, problem $P_1$ can be converted by:
	\begin{equation}
		\label{eq:30}
		\begin{array}{*{20}{l}}
			{{P_{3.3}}:}&{\mathop {\max }\limits_{\mathbf{w}} }&{2\Re \left\{ {{\boldsymbol{\zeta} ^H}\mathbf{w}} \right\} - {\mathbf{w}^H}\mathbf{L}\mathbf{w}},\\
			{}&{\mathrm{s.t.}}&{\eqref{eq:21C}},\end{array}
	\end{equation}
	where ${\boldsymbol{\zeta}}  = {{\big[\boldsymbol{\zeta}_1^T,\cdots,\boldsymbol{\zeta}_K^T \big]}^T}$, ${{\boldsymbol{\zeta}} _k^H = \sqrt {\varpi_k\big( {1 + {\varsigma _k}} \big)} \xi _k^*\mathbf{h}_k^H}$ and $\mathbf{L} = {\mathbf{I}_K} \otimes \big( {\sum\nolimits_{k = 1}^K {{{\left| {{\xi _k}} \right|}^2}{\mathbf{h}_k}\mathbf{h}_k^H} } \big)$.
	
	Since problem $P_{3.3}$ in \eqref{eq:30} is a standard quadratic constraint quadratic programming (QCQP) problem, by adopting the Lagrange multiplier method [22], the optimal $\mathbf{w}_k$ can be obtained by:
	\begin{equation}
		\label{eq:31}
		{\mathbf{w}^{opt}} = {\big( {\mathbf{L} + \lambda {\mathbf{I}_{KM}}} \big)^{ - 1}}{\boldsymbol{\zeta}},
	\end{equation}
	where $\lambda$ is the Lagrange multiplier, which should be chosen such that the complementary slackness condition of power constrain in $\eqref{eq:21C}$ is satisfied and the optimal Lagrange multiplier $\lambda$ can be obtained by grid search \cite{dean1999design}.
	
	\paragraph{Fix $\left( {\mathbf{w}, \boldsymbol{\varsigma}, \boldsymbol{\xi}} \right)$ and then optimize $\hat{\mathbf{v}}$}
	
	Recalling the superimposed channel ${\mathbf{h}}^H_k = {\hat{\mathbf{v}}^{H}\hat{\mathbf{C}}_k + \mathbf{h}_{bu,k}^H}$, and fixing ${\mathbf{w}, \boldsymbol{\varsigma}}$ and $\boldsymbol{\xi}$, problem $P_1$ can be converted by:
	\begin{equation}
		\label{eq:32}
		\begin{array}{*{20}{l}}
			{{P_{3.4}}:}&{\mathop {\max }\limits_{\hat{\mathbf{v}}} }&{ - {{\hat{\mathbf{v}}}^H}\mathbf{U}\hat{\mathbf{v}} - 2\Re \left\{ {{{\hat{\mathbf{v}}}^H}{\boldsymbol{\phi}}} \right\}},\\
			{}&{\mathrm{s.t.}}&{\eqref{eq:21D}},
		\end{array}
	\end{equation}
	Where ${\boldsymbol{\phi}} = \sum\nolimits_{k = 1}^K {{{\left| {{\xi _k}} \right|}^2}{{\hat{\mathbf{C}}}_k}\big( {\sum\nolimits_{j = 1}^K {{\mathbf{w}_j}\mathbf{w}_j^H} } \big){\mathbf{h}_{bu,k}}}  - \sum\nolimits_{k = 1}^K {\sqrt {\varpi_k\big( {1 + {\varsigma _k}} \big)} \xi _k^*{{\hat{\mathbf{C}}}_k}{\mathbf{w}_k}} $ and $\mathbf{U} = \sum\nolimits_{k = 1}^K {{{\left| {{\xi _k}} \right|}^2}{{\hat{\mathbf{C}}}_k}\sum\nolimits_{j = 1}^K {{\mathbf{w}_j}\mathbf{w}_j^H} \hat{\mathbf{C}}_k^H} $.
	
	Consequently, let ${\hat{\mathbf{v}}^t}$ denote the solution of the subproblem at the $t$-th iteration. Recalling Lemma 1 from \cite{7362231}, the objective function of problem $P_{3.4}$ can be approximate by constructing a series of more tractable subproblems as follows:
	\begin{equation}
		\label{eq:33}
		{{\hat{\mathbf{v}}}^H}\mathbf{U}\hat{\mathbf{v}} \le {{\hat{\mathbf{v}}}^H}\mathbf{X}\hat{\mathbf{v}} - 2\Re\left \{ {{\hat{\mathbf{v}}}^H}\left ( \mathbf{X} - \mathbf{U} \right ) \hat{\mathbf{v}}^t \right \} + {\left ( \hat{\mathbf{v}}^t \right ) }^H\left ( \mathbf{X} - \mathbf{U} \right ) \hat{\mathbf{v}}^t,
	\end{equation}
	where $\mathbf{X} = \lambda_{\max }\mathbf{I}_Q$ and $\lambda_{\max }$ is the maximum eigenvalue of $\mathbf{U}$.
	
	Therefore, the subproblem to be solved at the $t$-th iteration is given by
	\begin{equation}
		\label{eq:34}
		\begin{array}{*{20}{l}}
			{{P_{3.5}}:}&{\mathop {\max }\limits_{\hat{\mathbf{v}}} }&{2\Re \left\{ {{{\hat{\mathbf{v}}}^H}{\boldsymbol{\varphi} ^t}} \right\}},\\
			{}&{\mathrm{s.t.}}&{\eqref{eq:21D}},
		\end{array}
	\end{equation}
	where ${\boldsymbol{\varphi} ^t} = \big( {{\lambda _{\max }}{\mathbf{I}_Q} - \mathbf{U}} \big){\hat{\mathbf{v}}^t} - \boldsymbol{\phi} $. 

	The optimal solution of problem $P_{1.4}$ is given by
	\begin{equation}
		\label{eq:35}
		{\hat{\mathbf{v}}^{t + 1}} = {e^{j\mathrm{arg}\big( {{\boldsymbol{\varphi} ^t}} \big)}},
	\end{equation}
	
	Finally, for clarity, we summarize the proposed two-stage algorithm scheme in Algorithm \ref{alg:1}.
	\begin{algorithm}[htbp]
		\caption{Proposed Two-Stage Algorithm Scheme}
		\label{alg:1}
		\renewcommand{\algorithmicrequire}{\textbf{Input:}}
		\renewcommand{\algorithmicensure}{\textbf{Output:}}
		\begin{algorithmic}[1]
			\REQUIRE Channels $\hat{\mathbf{C}}_k$, $\mathbf{h}_{bu,k}$, $\bar{\mathbf{C}}_k$ and $\bar{\mathbf{h}}_{bu,k}$, $\forall k \in \left\{ {1, \cdots ,K} \right\}$  
			\ENSURE Optimized BS beamforming vector $\mathbf{w}$, optimized IEG-IRS combination-reflection precoding vector $\hat{\mathbf{v}}$, optimized grouping strategy matrix $\mathbf{G}$ and maximized WSR $R$  
			
			\STATE  Initialize $\mathbf{w}$, $\hat{\mathbf{v}}$ and $\mathbf{G}$ randomly
			
			\STATE Optimized grouping strategy matrix $\mathbf{G}$ by \eqref{eq:24}
			
			\WHILE{no convergence of $R$}
			\STATE Update $\boldsymbol{\varsigma}$ and $\boldsymbol{\xi}$ by \eqref{eq:29};
			\STATE Update $\mathbf{w}$ by \eqref{eq:31};
			\STATE Update $\hat{\mathbf{v}}$ by \eqref{eq:35};
			\ENDWHILE
			
			\RETURN Optimized $\mathbf{w}$, $\hat{\mathbf{v}}$, $\mathbf{G}$ and $R$
		\end{algorithmic}
	\end{algorithm}
	
	\section{Simulation Results}
	In this section, simulation results are present to validate the effectiveness of the proposed Algorithms \ref{alg:1}. Specifically, in Subsection V-A, the simulation setups are provided to build an IEG-IRS aided MU-MISO system model. Then, in the following subsections, we consider the four comparative schemes to WSR, including the number of groups, the number of IRS elements, distance and power for simulations.
	
	\subsection{Simulation setup}
	\begin{figure}[htbp]
		\centerline{\includegraphics[width=\columnwidth]{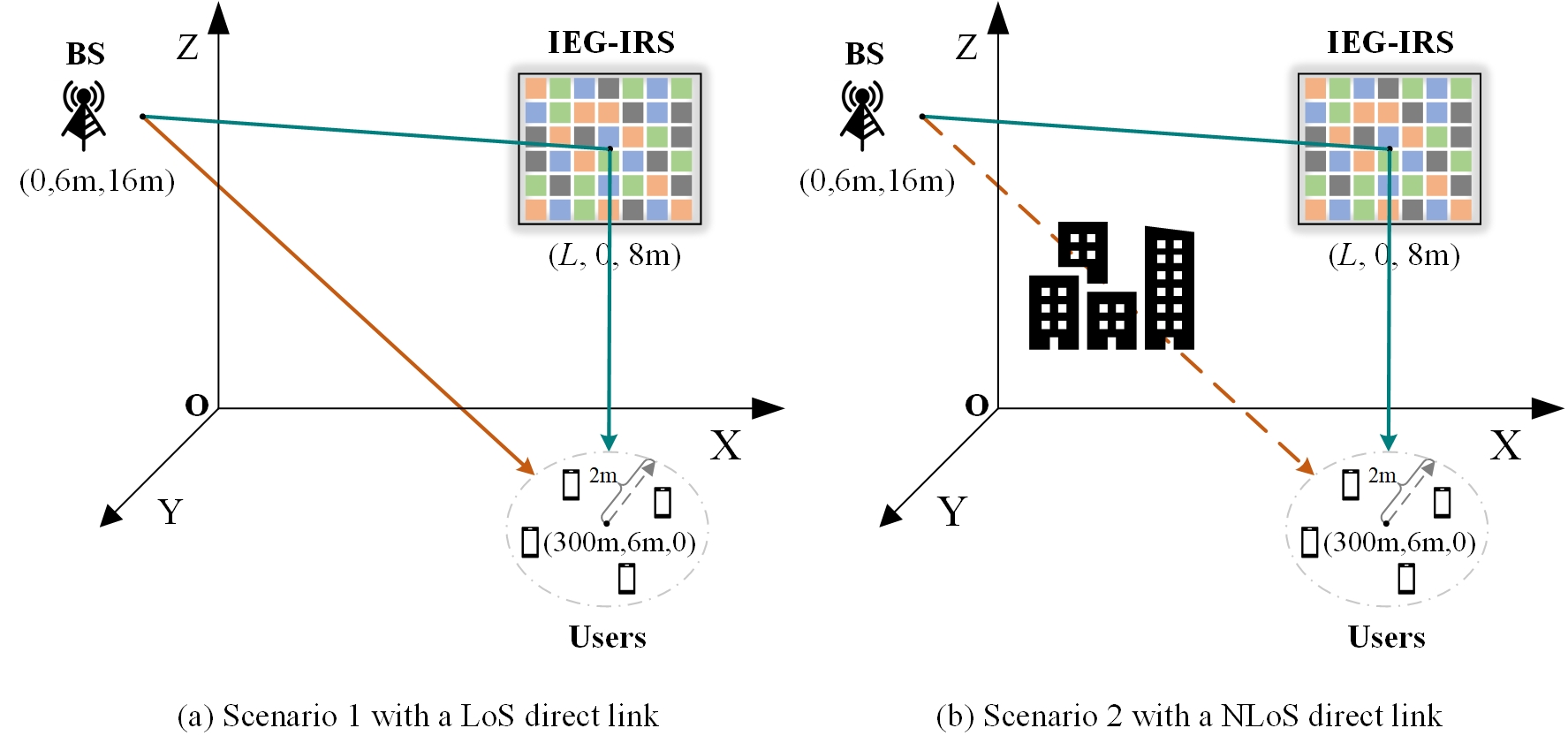}}
		\caption{An illustration of a BS aided by an IEG-RIS serves four users.}
		\label{fig:3}
	\end{figure}
	
	As illustrated in Fig.\ref{fig:3}, we consider an IRS aided MU-MISO system operating at a frequency of 5 GHz. Particularly, two typical scenarios with different channel conditions are considered, where the direct link is severely obscured by obstacles in Fig.\ref{fig:3} (a), while the direct link is unobscured in Fig.\ref{fig:3} (b). Then, for the large-scale fading, two different path loss models from the 3GPP standard \cite{3gpp36814} are adapted as
	\begin{equation}
		\label{eq:36}
		\left\{ {\begin{array}{*{20}{l}}
				{P{L_{los}} = 42.0 + 22.0\log d}, \\
				{P{L_{nlos}} = 40.9 + 36.7\log d},
			\end{array}} \right.
	\end{equation}
	where $d$ denotes the individual link distance between two devices. For the two considered scenarios, path loss model $PL_{los}$ is used to model the BS-IRS link, the IRS-user link and the unobscured BS-user link, and $PL_{nlos}$ is used to generate the obscured BS-user link, respectively.
	
	As common settings, the numbers of users and BS antennas are set as $K=4$ and $M = 4$, respectively. The BS equiped with ULA is located at $(0, 6\mathrm{m}, 16\mathrm{m})$ and the four users are randomly located in a ball with a radius of $2$m from the center $\big(300\mathrm{m}, 6\mathrm{m}, 0\mathrm{m}\big)$, respectively. Unless specified otherwise, the number of the IRS elements is set to $N=10000$ and the IRS equiped with uniform planar array (UPA) is located at $(300\mathrm{m}, 0, 8\mathrm{m})$. Recalling \eqref{eq:4}, we assume $\kappa_{bi} = \kappa_{iu} = \kappa_{bu} = 1$ for the BS-IRS link, the IRS-user link and the BS-user link, respectively, and $\sigma^2=-100\mathrm{dBm}$ for all users. For fair comparison, the transmit power consumption is constrained by $P_{\mathrm{max}} = 10\mathrm{dBm}$ for the other benchmark systems.
	
	Then, to show the effectiveness of the proposed joint design, we compare the following four benchmarks for simulations:
	\begin{itemize}
		\item IEG-IRS: In an IEG-IRS aided MU-MISO system, the proposed Algorithm 1 is employed to optimize the beam-domain grouping strategy and the multi-user active-passive beamforming;
		\item AEG-IRS: In an AEG-IRS aided MU-MISO system, the algorithm proposed in \cite{9039554} is adopted to jointly optimize transmit beamforming at BS and adjacent-grouping passive precoding at AEG-IRS;
		\item U-IRS: In a U-IRS aided MU-MISO system, the algorithm proposed in \cite{9090356} is utilized to jointly optimize active beamforming and constrained passive precoding, treating the remaining uncontrolled reflective links as combined channels from the BS to the users;
		\item Random RCV: In an IRS-aided MU-MISO system, the RCV of all IRS elements is randomly set. Thus, the combined channels from the BS to users are estimated and the weighted mean-squared error minimization (WMMSE) algorithm proposed in \cite{5756489} is used to optimize the BS beamforming;
		\item Without IRS: Similar as random RCV, the WMMSE algorithm from \cite{5756489} is adopted to optimize the BS beamforming.
	\end{itemize}
	
	\subsection{WSR versus number of groups}
\begin{figure}[htbp]
		\centerline{\includegraphics[width=\columnwidth]{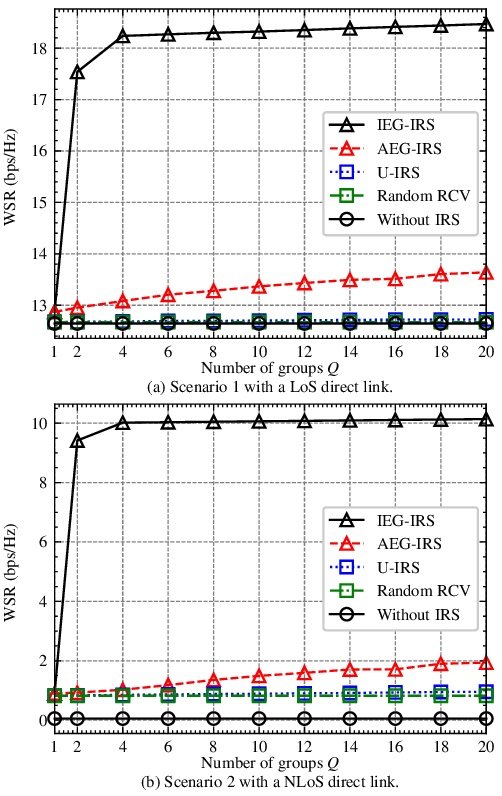}}
		\caption{Simulation results for the WSR versus number of groups $Q$ in an IEG-IRS aided MU-MISO system ($N=10000$).}
		\label{fig:4}
	\end{figure}

	To observe the effect of different grouping numbers on the WSR, we assume that a $10000$-element IRS is adopted. Then, in Fig. \ref{fig:4} (a) and Fig. \ref{fig:4} (b), we plot the users' WSR versus the number of IEG-IRS groups for the two considered scenarios. From Fig. \ref{fig:4}, it is observed that as the number of groups ranges from $1$ to $2$, the system capacity achieved by the IEG-IRS has significantly increased. Subsequently, while the growth rate of system capacity has decelerated, it continues to maintain an upward trajectory with an increasing number of groups. This is because when the number of IRS elements in each group is sufficiently large, the system capacity loss caused by a small number of groups ($Q > 1$) can be ignored from \eqref{eq:18}. Besides, the other benchmarks only achieve a limited WSR gain, while our proposed IEG-IRS achieves a much higher WSR gains. For example, when $Q=4$, the WSR aided by without IRS, AEG-IRS and IEG-IRS in scenario 1 are 12.65 $\mathrm{bps/Hz}$, 13.08 $\mathrm{bps/Hz}$ and 18.24 $\mathrm{bps/Hz}$, respectively, while in scenario 2, the values are 0.05 $\mathrm{bps/Hz}$, 1.02 $\mathrm{bps/Hz}$ and 10.02 $\mathrm{bps/Hz}$, respectively. For those results, the AEG-IRS obtains a 3 $\%$ gain in scenario 1 and a 1940 $\%$ gain in scenario 2. In contrast, the IEG-IRS achieves a much higher WSR gains of 44 $\%$ in scenario 1 and 19940 $\%$ in scenario 2 than others. It demonstrates that, compared with other benchmarks, the implementation of our proposed grouping strategy has the potential to facilitate enhanced communication for an XL-IRS in low pilot overhead conditions. Besides, when the number of groups is small ($Q > 1$), IEG-IRS can provide significant performance gains. In the subsequent simulation experiments, the number of the groups will be set by $Q=4$.
	
	\subsection{WSR versus number of IRS elements N}
\begin{figure}[htbp]
		\centerline{\includegraphics[width=\columnwidth]{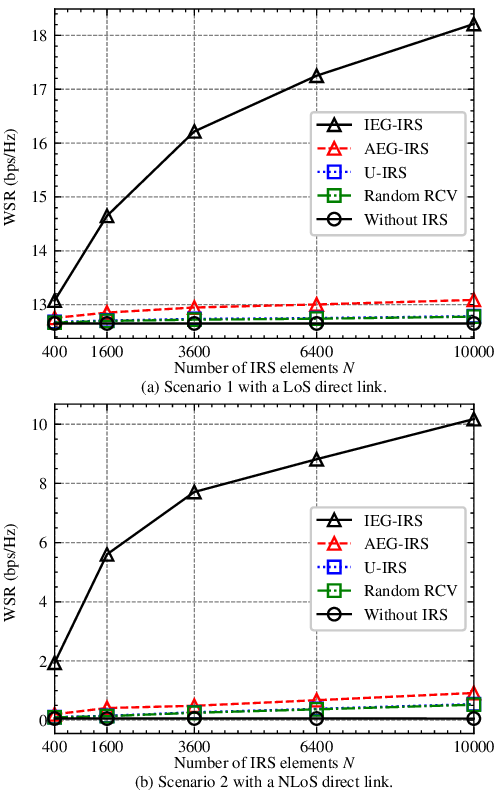}}
		\caption{Simulation results for the WSR versus number of IRS elements $N$ in an IEG-IRS aided MU-MISO system ($Q=4$).}
		\label{fig:5}
	\end{figure}

	For the same setup as in Fig. \ref{fig:4}, we plot the users' WSR versus the number of IRS elements $N$ with the number of groups $Q=4$ in Fig. \ref{fig:5} for the two considered scenarios. The optimal grouping strategy has been determined by the deterministic S-CSI in advance, and the real-time pilot overhead required for IEG-IRS are consistent with those of other benchmarks. As the number of IRS elements $N$ increases, it is observed that the IEG-IRS achieves higher and more sustainable system capacity than the other benchmarks, and the performance improvement for the IEG-IRS aided communication system is much larger than that of other benchmarks. For example, when $N$ increases from 400 to 10000, the WSR of the AEG-IRS aided system increases from 12.75 $\mathrm{bps/Hz}$ to 13.09 $\mathrm{bps/Hz}$ in scenario 1 and increases from 0.20 $\mathrm{bps/Hz}$ to 0.92 $\mathrm{bps/Hz}$ in scenario 2, respectively. By contrast, the WSR of the IEG-IRS aided system increases from 13.08 $\mathrm{bps/Hz}$ to 18.21 $\mathrm{bps/Hz}$ in scenario 1 and increases from 1.94 $\mathrm{bps/Hz}$ to 10.17 $\mathrm{bps/Hz}$ in scenario 2, respectively. These results show that, compared with the other benchmarks, increasing the number of IRS elements of IEG-IRS is much more efficient for improving the system capacity, which is consistent with the performance analysis described in Section III.
	
	\subsection{WSR versus AP-IRS distance}
\begin{figure}[htbp]
		\centerline{\includegraphics[width=\columnwidth]{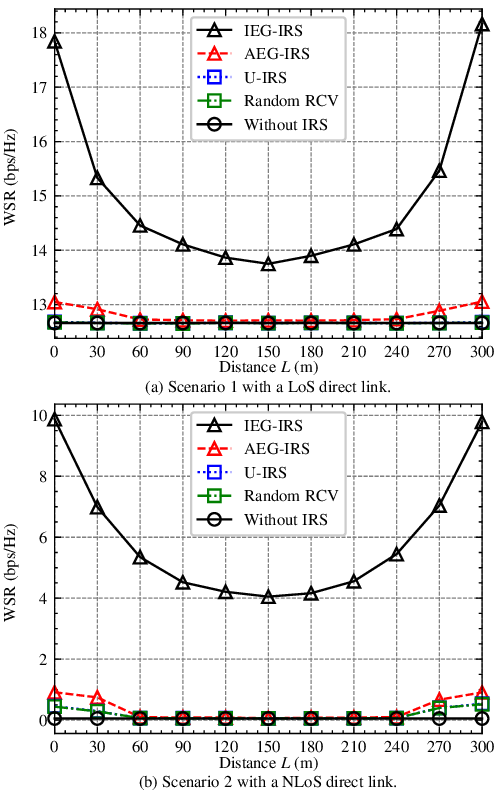}}
		\caption{Simulation results for the WSR versus AP-IRS distance $L$ in an IEG-IRS aided MU-MISO system ($N=10000$ and $Q=4$).}
		\label{fig:6}
	\end{figure}

	Same as the setups in Subsection V-B, we plot the WSR versus AP-IRS distance $L$ to show the notable performance improvement by the IEG-IRS for two considered scenarios in Fig. \ref{fig:6}. It is assumed that the IRS is located at $(L, 0, 8\mathrm{m})$, where $L$ ranges from 0 to $300\mathrm{m}$. And a $10000$-element IEG-IRS and $Q=4$ are adopted. From Fig. \ref{fig:6}, it is observed that across all distance ranges, the WSR achieved by the IEG-IRS is significantly higher than that of other benchmarks. Further, even when the IRS is located in the middle of the BS and the users, the performance improvement for the IEG-IRS-aided communication system is visible. For example, when $L=150\mathrm{m}$, the WSR aided by without IRS, AEG-IRS and IEG-IRS in scenario 1 are 12.65 $\mathrm{bps/Hz}$, 12.71 $\mathrm{bps/Hz}$ and 13.75 $\mathrm{bps/Hz}$, respectively, while in scenario 2, the values are 0.05 $\mathrm{bps/Hz}$, 0.06 $\mathrm{bps/Hz}$ and 4.05 $\mathrm{bps/Hz}$, respectively. From these results, it shows that for an XL-IRS, our proposed IEG-IRS is promising for overcoming ``multiplicative fading'' and improving system capacity in low pilot overhead conditions even when the direct link is unobscured.
	
	\subsection{WSR versus BS transmit power}
\begin{figure}[htbp]
		\centerline{\includegraphics[width=\columnwidth]{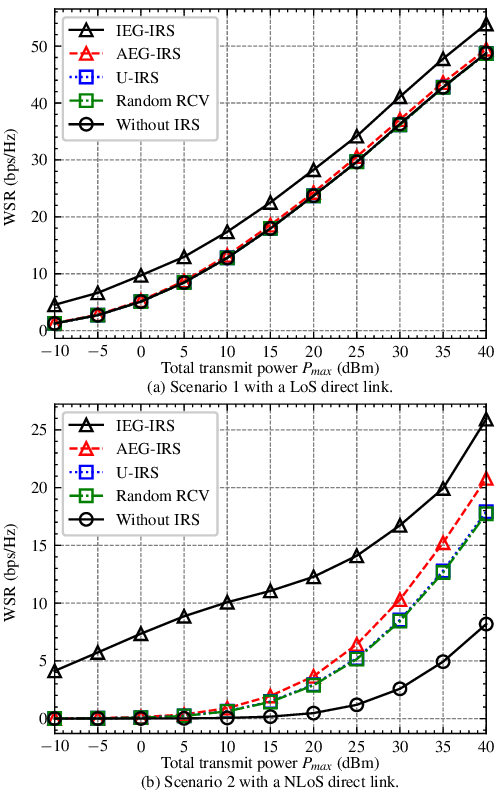}}
		\caption{Simulation results for the WSR versus BS transmit power $P_{\mathrm{max}}$ in an IEG-IRS aided MU-MISO system ($N=10000$ and $Q=4$).}
		\label{fig:7}
	\end{figure}

	To evaluate the averaged performance in the coverage of IEG-RIS, same as the setups in Fig. \ref{fig:4}, the users' WSR versus the total power consumption $P_{\mathrm{max}}$ is shown in Fig. \ref{fig:7}. It is observed that the required power consumption for the IEG-IRS aided system is much lower than that of other benchmarks in both scenarios. For example, when the total power consumption of the AEG-IRS aided system is $P_{\mathrm{max}}=25$ dBm, the IEG-IRS aided system only requires 21.5 dBm in scenario 1 and requires -2.5 dBm in scenario 2 for achieving the same WSR. The reason behind for this result is that, the impact of the large ``multiplicative fading'' on IEG-IRS is much weaker than that of other benchmarks. Thus, the IEG-IRS is promising for improving the energy efficiency of communication systems.

	\section{Conclusion}
	In this paper, we have analyzed the characteristics of XL-IRS and proposed the concept of IEG-IRS to overcome the fundamental bottlenecks between high computational complexity and pilot overhead. Specifically, we have presented the signal model for the IEG-IRS and given the intelligent grouping strategy for reducing pilot overhead. Based on the proposed signal model, we have analyzed the asymptotic performance gain for IEG-IRS and then the problem of WSR maximization in an IEG-IRS-aided MU-MISO system is formulated. Subsequently, a two-stage algorithm design is proposed to solve this problem. Finally, simulation results have shown that, compared with the other benchmarks realizing only a limited WSR gain, while the proposed IEG-RIS can achieve a significant WSR gain in a typical communication system, thus indeed overcoming the performance drawbacks. Moreover, it is not difficult to observe that our proposed grouping strategy can also be applied to multiple IRSs and help to solve the high pilot overhead caused by the introduction of multiple IRSs. As such, in the future, many researches on XL-IRS are worth pursuing through the proposed IEG-RIS, including multi-IRSs assistance \cite{9893192}, channel estimation \cite{9866003} and near-field modeling \cite{10287779}.
	
	\appendices
	\section{PROOF OF Lemma 1}
	Recalling that $\mathbf{c} = \mathbf{c}_1 + \mathbf{c}_2$, where $\mathbf{c}_1 = {\bar{a} {\delta_{bi}}{\delta_{iu}} \bar{\mathbf{h}}_{iu}^* \odot \bar{\mathbf{h}}_{bi}^*}$ and $\mathbf{c}_2 = {
		\tilde{a}{\delta_{bi}}{\delta_{iu}} \tilde{\mathbf{h}}_{iu}^* \odot \tilde{\mathbf{h}}_{bi}^*
		+ \bar{b}{\delta_{bi}}{\delta_{iu}} \bar{\mathbf{h}}_{iu}^* \odot \tilde{\mathbf{h}}_{bi}^* + \tilde{b}{\delta_{bi}}{\delta_{iu}} \tilde{\mathbf{h}}_{iu}^* \odot \bar{\mathbf{h}}_{bi}^*
	}$ denote the cascaded deterministic and stochastic components, respectively. It is easy to verify that the cascaded deterministic and stochastic components are statistically independent. For the combined cascaded stochastic component denoted by $\hat{\mathbf{c}}_2 = \mathbf{G}\mathbf{c}_2$, by invoking the Lindeberg-L$\mathrm{\acute{e}}$vy central limit theorem,  we have $\hat{\mathbf{c}}_2 \sim \mathcal{CN}\big(0,\mu \delta_{bi}^2\delta_{iu}^2\big( {1 - \bar{a}^2}\big){\mathbf{I}_Q} \big)$ with any grouping, which each group contains the same number of IRS elements. For the combined cascaded deterministic component denoted by $\hat{\mathbf{c}}_1 = \mathbf{G}\mathbf{c}_1$, we have ${{\mathbf{c}}_1} = {\big[ c_{1,1},\cdots,c_{1,n},\cdots,c_{1,N} \big]^T}$ from \eqref{eq:9} and \eqref{eq:10}, where $c_{1,n} = \bar a{\delta_{bi}}{\delta_{iu}}e^{{-j2\pi\left( {n - 1} \right)\Delta }}$ and $\Delta = \frac{{\sin {\theta_{bi}} + \sin {\theta_{iu}}}}{2} \in \big[ { 0, 1} \big)$. Then, the phases of the $N$-element response vector is preprocessed as
	\begin{equation}
		\label{eq:37}
		\left\langle {\tilde{\boldsymbol{\Theta}}} \right\rangle  = \tilde{\boldsymbol{\Theta}} - \left\lfloor {\tilde{\boldsymbol{\Theta}}} \right\rfloor,
	\end{equation}
	where ${\tilde{\boldsymbol{\Theta}}}  = \big[0,\cdots,\big( {N - 1} \big)\Delta \big]$, $\big\lfloor{\tilde{\boldsymbol{\Theta}}} \big\rfloor$ denotes the integer part of ${\tilde{\boldsymbol{\Theta}}}$ and ${{\big\langle {\tilde \Theta } \big\rangle }_i} \in \big({ 0,1} \big)$.
	
	Subsequently, the two cases of $\Delta$ as an irrational number and a rational number are further analyzed. Let $\hat{\mathbf{c}}^i_1$ and $\hat{\mathbf{c}}^r_1$ denote the combined cascaded deterministic component $\hat{\mathbf{c}}$ in the cases of irrational and rational numbers, respectively.
	
	Firstly, when $\Delta$ is an irrational number, we assume that $\big\langle{\tilde{\boldsymbol{\Theta}}} \big\rangle$ is sorted in ascending order. For simplicity, recalled that $1,\cdots,N$ denotes the reordered indexs. By invoking the Weyl's equidistribution theorem \cite{stein2011fourier} and the phase similarity clustering method, we have ${{\hat{c}}^{i}_{1,q}} = {\delta_{bi}}{\delta_{iu}}\bar{a}\sum\nolimits_{i = \left( {q - 1} \right)\mu  + 1}^{q\mu } {{e^{-j2\pi {{\left\langle {\tilde \Theta } \right\rangle }_i}}}}  \to N{\delta_{bi}}{\delta_{iu}}\bar{a}\int_{\frac{{q - 1}}{Q}}^{\frac{q}{Q}} {{e^{-j2\pi \theta }}} d\theta \to \mu {\delta_{bi}}{\delta_{iu}}\imath\bar{a}{e^{-j\frac{{\left( {2q - 1} \right)\pi }}{Q}}}$ for $\forall q$ as $\mu \to \infty$, where $\hat{c}^{i}_{1,q}$ is the $q$-th elements in $\hat{\mathbf{c}}^r_1$ and $\imath = {\frac{{\sin \frac{\pi }{Q}}}{{\frac{\pi }{Q}}}}$. Then, by using the fact, it follows that
	\begin{equation}
		\label{eq:38}
		\hat{\mathbf{c}}_1^{i} \to \mu{\delta_{bi}}{\delta_{iu}}\imath\bar{a}\boldsymbol{\vartheta},
	\end{equation}
	where $\boldsymbol{\vartheta} = {\big[
			e^{-j\frac{{\pi }}{Q}},\cdots,e^{-j\frac{{\left( {2Q - 1} \right)\pi }}{Q}}\big ]^H }$ and $\imath = {\frac{{\sin \frac{\pi }{Q}}}{{\frac{\pi }{Q}}}}$.
	
	Secondly, when $\Delta$ is a rational number, a straightforward idea is that it is always possible to find an irrational number that approximates a given rational number. By using the process of proof for the above irrational case, the rational case completes the proof. A more rigorous proof is presented below.
	
	Assuming that a fraction with a large denominator is approximate to $\Delta$, where the denominator satisfies both a factor of $N$ and a multiple of $Q$, we then have that $\Delta  \approx \frac{x}{y}$, $T  = \frac{N}{y}$ and $\eta  = \frac{y}{Q}$, where $x$, $y$, $T$ and $\eta$ are integer, and the value of $y$ is as large as possible. It means that the $N$ elements in ${\big\langle {\tilde{\boldsymbol{\Theta}}} \big\rangle }$ can be divided into $T$ identical arrays denoted by $\mathbf{Y} = \big[ 1,\cdots,\frac{y-1}{y} \big]$ and $\mu=\eta T$. Then, by using the phase similarity clustering method, we have ${\hat c^{r}_{1,q}} \approx T{\delta_{bi}}{\delta_{iu}}{e^{-j2\pi \frac{{\left( {q - 1} \right)\eta }}{y}}}\sum\nolimits_{i = 0}^{\eta-1}  {{e^{-j2\pi \frac{i}{y}}}} \to \mu {\delta_{bi}}{\delta_{iu}}\imath\bar{a}{e^{-j\pi\frac{{\left( {2q\eta-\eta - 1} \right) }}{Q\eta}}} \approx \mu {\delta_{bi}}{\delta_{iu}}\imath\bar{a}{e^{-j\frac{{\left( {2q - 1} \right)\pi }}{Q}}}$ for $\forall q$ as $\eta \to \infty$, where $\hat{c}^{r}_{1,q}$ is the $q$-th elements in $\hat{\mathbf{c}}^{r}_1$. By using the fact, it follows that
	\begin{equation}
		\label{eq:39}
		\hat{\mathbf{c}}_1^{r} \to \mu{\delta_{bi}}{\delta_{iu}}\imath\bar{a}\boldsymbol{\vartheta},
	\end{equation}
	
	As a consequence, we have $\hat{\mathbf{c}}_1 \to \mu{\delta_{bi}}{\delta_{iu}}\imath\bar{a}\boldsymbol{\vartheta}$. Recalling that $\mathbf{c} = \mathbf{c}_1 + \mathbf{c}_2$, the distribution for combined cascaded channel $\hat{\mathbf{c}}$ is given by
	
	\begin{equation}
		\label{eq:40}
		\hat{\mathbf{c}} \sim \mathcal{CN}\big( \mu{\delta_{bi}}{\delta_{iu}}\imath \bar{a}\boldsymbol{\vartheta} ,\mu \delta_{bi}^2\delta_{iu}^2\big( {1 - \bar{a}^2}\big){\mathbf{I}_Q} \big),
	\end{equation}
	This completes the proof.
	
	\bibliographystyle{IEEEtran}
	\bibliography{IEG-IRS}
	
\end{document}